\documentclass[aps,prb,twocolumn,superscriptaddress,longbibliography]{revtex4-2}
\usepackage{amsmath,amssymb}
\usepackage[pdftex]{hyperref,graphicx}
\hypersetup{colorlinks = true, urlcolor = blue, linkcolor = blue, citecolor = blue}
\usepackage{bm}
\usepackage[normalem]{ulem}

\begin{document}
\title{Disorder-driven loss of crystalline coherence in two-dimensional electron solids: Quantum Hartree--Fock and classical molecular dynamics}
\author{Sankar Das Sarma}
\affiliation{Condensed Matter Theory Center and Joint Quantum Institute, Department of Physics, University of Maryland, College Park, Maryland 20742, USA}

\author{Haining Pan}
\affiliation{Department of Physics, University of Florida, Gainesville, Florida 32611, USA}

\begin{abstract}
We investigate the effects of quenched disorder on two-dimensional electron solids using Hartree--Fock theory and classical molecular dynamics. Increasing disorder weakens crystalline correlations, broadens the structure factor, and reduces the measured coherence length toward the pristine lattice spacing. This evolution is consistent with the Imry--Ma and Larkin picture: random pinning disrupts translational order on sufficiently long scales, and stronger disorder reduces the distance over which crystalline organization survives. Classical simulations further show that stronger pinning raises the temperature for particle rearrangement even after crystalline order has disappeared. Disorder thus suppresses structural coherence while stabilizing an amorphous solid against thermal rearrangement.
\end{abstract}
\maketitle
\section{Introduction}\label{sec:introduction}

How does random pinning destroy the spatial order of a two-dimensional electron solid? This question applies equally to quantum and classical Wigner crystals. In both systems, Coulomb repulsion favors a triangular arrangement, while randomly positioned impurities favor a different, spatially irregular density pattern. The clean quantum crystal competes with kinetic energy and exchange; the classical crystal competes with thermal motion~\cite{wigner1934interaction,grimes1979evidence}. Once impurities are present, the central structural question is how far crystalline organization survives and how that distance decreases with increasing disorder.

This paper contains two parallel studies of that problem. The quantum study uses continuum Hartree--Fock (HF) theory to follow the self-consistent electron density in a short-range random potential. The classical study uses molecular dynamics (MD) and zero-temperature relaxation to follow Coulomb particles in randomly positioned attractive wells. Each calculation has its own Hamiltonian, clean reference state, disorder control parameter, and numerical checks. We compare their density correlations, scattering intensities, and structural crossover diagrams on equal footing. The classical calculation additionally determines how disorder changes the temperature scale for particle rearrangement.

For the quantum problem, correlated calculations place clean crystallization at much lower densities than HF. Tanatar and Ceperley reported $r_s=37\pm5$~\cite{tanatar1989ground}, Drummond and Needs found a paramagnetic-fluid to antiferromagnetic-crystal transition at $r_s=31(1)$~\cite{drummond2009phase}, and Azadi, Drummond, and Vinko obtained $r_s=35(1)$~\cite{azadi2024quantum}. The hybrid phase proposed by Falakshahi and Waintal was not found in the search of Drummond and Needs~\cite{drummond2009phase,falakshahi2005hybrid}. Unrestricted HF instead supports crystalline states at much smaller $r_s$~\cite{trail2003unrestricted,bernu2011hartree}. Our quantum calculation uses this mean-field solid to study the competition between electronic order and impurities. A direct-only Hartree calculation with exactly the same disorder and discretization supplies a control that isolates the role of exchange. The advantage of using the mean field HF technique is that the physics is transparent, even if the actual $r_s$ regime of interest shifts to lower $r_s$, which is not a problem for the qualitative understanding of the subtle competition between interaction and disorder of central interest to the current work. We mention in this context that even the fully numerical correlated Monte Carlo works give different critical $r_s$ for crystallization depending on the specific calculations as mentioned above.

The classical problem addresses the same competition without orbital kinetic energy or exchange. Classical electron crystallization has an experimental realization on liquid helium~\cite{grimes1979evidence}, and simulations of the clean two-dimensional Coulomb system provide melting benchmarks~\cite{mazars2015melting}. Quenched-pin simulations also show that disorder can change both spatial organization and thermal behavior~\cite{reichhardt2023melting}. Our classical calculation resolves the particle configurations, coordination defects, positional and orientational correlations, and cage escape in an open neutralized square. Varying both impurity density and electron density gives a structural map; heating at fixed impurity configurations separates loss of crystal-like order from thermal rearrangement of an already amorphous solid.

Impurity effects on the quantum fluid--solid competition~\cite{chui1995impurity} and collective pinning of an electron crystal~\cite{chitra2005zero} provide the broader setting. The Imry--Ma random-field argument and the Larkin elastic picture explain why the pristine translationally ordered state is unstable on sufficiently long scales under arbitrarily weak random pinning~\cite{imry1975random,larkin1970effect}. Thus a finite-disorder line separating a perfect crystal from an amorphous solid is not the appropriate organizing principle here. The relevant evolution is a crossover: disorder reduces the spatial range and strength of crystalline correlations. Weak disorder can leave a finite field of view almost indistinguishable from a crystal, whereas stronger disorder brings the residual coherence down to the nearest-neighbor scale. We connect this physical picture to explicit correlation measures and model-specific elastic estimates below.

Recent imaging of two-dimensional electron and hole solids makes these distinctions directly relevant. Xiang and co-workers reported disorder-pinned hole nanocrystals in bilayer MoSe$_2$~\cite{xiang2025imaging}. Ge and co-workers combined electron-density imaging with neural-quantum-state calculations to study quenched disorder~\cite{ge2026visualizing}. Babbar, Li, and Das Sarma argued that strongly disordered patterns discussed in the earlier Ge preprint can instead represent an impurity-localized Anderson solid~\cite{babbar2026wigner}. A very recent imaging experiment from Princeton on graphene~\cite{ali} concludes that disorder suppresses crystallization leading to amorphous solid structure as discussed in Ref.~\cite{ge2026visualizing}. Large density contrast alone cannot resolve this issue: both a crystal and an amorphous pinned assembly can have strongly modulated density. Spatial correlations and their Fourier spectra test crystalline organization, while mobility measurements test rearrangement. The present work evaluates the former in both quantum and classical models and the latter through classical dynamics.

The two calculations use complementary short-range disorder protocols. HF increases the amplitude of one fixed Gaussian random potential. MD increases the concentration of attractive wells in one nested Poisson impurity field while holding their physical depth and radius fixed. Neither calculation averages over disorder realizations. Their disorder scales are therefore defined separately, and the comparison concerns a common physical mechanism rather than a numerical conversion between the two disorder axes. Disorder averaging is inappropriate in the current context as experiments are looking at specific disorder configurations in specific samples through direct imaging.

Our main result is a quantitative account of the loss of crystalline coherence under increasing random pinning in both descriptions, which is anticipated qualitatively in Refs.~\cite{reichhardt2023melting,ali,babbar2026wigner,huang2026interplay}. The HF elastic density spectrum transfers weight away from the pristine Bragg vectors; the classical structure factor evolves from sharp sixfold peaks into diffuse scattering. A common first-shell density-correlation construction quantifies this evolution. Its resolved strong-disorder scale is of order one pristine lattice spacing, while weak-disorder values exceed the available observation range. The classical system simultaneously becomes harder to rearrange thermally. A disordered solid can therefore become more strongly pinned while becoming less crystalline. The new elements here are the continuum HF supercell with a same-disorder direct-only Hartree control, the HF Bragg fraction falling while the density contrast rises, the HF shear modulus entering a quantum Larkin estimate, the classical two-density map at fixed well parameters, the rise of the cage-escape scale with pin concentration, and one coherence diagnostic applied to both models.

Sections~\ref{sec:quantum_model}--\ref{sec:hf_crossover} present the quantum model, its clean and disordered results, and its structural results. Sections~\ref{sec:classical_model}--\ref{sec:thermal} give the classical model, the two-density structural map, and the thermal crossover. Section~\ref{sec:coherence} compares the spectra and correlation lengths directly; Sec.~\ref{sec:larkin} derives weak-disorder Larkin estimates for both models. Section~\ref{sec:discussion} states the numerical scope and the distinction between finite-scale crystalline appearance and long-range order. The two calculations answer the same structural question through different microscopic descriptions.

\section{Quantum model and variational equations}\label{sec:quantum_model}

\subsection{Units and disorder}\label{sec:units}

Let $a_B^*$ and $H^*$ denote the effective Bohr radius and Hartree. We define
\begin{equation}\label{eq:density_EF}
    n = \frac{1}{\pi (r_s a_B^*)^2}, \qquad E_F = \frac{2H^*}{r_s^2}
\end{equation}
for a fully polarized gas, since $k_F^2 = 4\pi n$ and $E_F = \hbar^2 k_F^2/(2m^*)$. With $x = r/(r_s a_B^*)$, the Hamiltonian in effective Hartree units is
\begin{equation}\label{eq:H_quantum}
    \begin{split}
        H={}&-\frac{1}{2r_s^2}\sum_i\nabla_{x_i}^2
        +\frac{1}{r_s}\sum_{i<j}\frac{1}{|x_i-x_j|}\\
        &+\sum_i u(x_i)+H_{\mathrm{bg}}.
    \end{split}
\end{equation}
The neutralizing background removes the zero-wavevector direct contribution. The supercell contains $N_c = 9$ electrons and has scaled area $A_c = \pi N_c$. It spans three primitive lattice vectors in each direction of a triangular crystal. The pristine lattice constant is
\begin{equation}\label{eq:aWC}
    a_{\mathrm{WC}} = \left(\frac{2\pi}{\sqrt{3}}\right)^{1/2} r_s a_B^*.
\end{equation}
A 32 \(\times\) 32 grid in fractional supercell coordinates is populated with independent unit-normal Gaussian numbers using NumPy's default generator with seed 1701. The grid mean is subtracted, but its sample variance is not rescaled. If $z_d$ denotes its normalized discrete Fourier transform (the first array index runs along the second fractional coordinate), the potential matrix is
\begin{equation}\label{eq:U_disorder}
    U_{GG'} = \frac{2\eta}{r_s^2} z_{G-G'}, \qquad \eta = \frac{W}{E_F}.
\end{equation}
The sampled amplitudes are $\eta = 0,\allowbreak 0.5,\allowbreak 1,\allowbreak 2,\allowbreak 4$. The saved solver parameter is $w = 2\eta$, so that $W = wH^*/r_s^2$. The grid potential thus follows $\mathcal{N}(0,W^2)$. The same realization is used at every point; there is no disorder averaging. This defines a grid-regularized short-range potential projected into the orbital basis. The cell and disorder cutoff scale with $r_s$; varying $r_s$ at fixed $\eta$ therefore does not describe a density sweep in one fixed laboratory potential. Furthermore, the disorder cell is periodically repeated. The results concern its structure, not localization in an infinite independently disordered system.

\subsection{Density matrix and self-consistent fields}\label{sec:self_consistent}

The orbital basis is $\exp[i(k+G)\cdot x]/\sqrt{\Omega}$. The reciprocal primitive vectors have angle 60\(^{\circ}\) and determinant $4\pi/N_c$. For each of $N_k = M^2$ midpoint Bloch momenta, define
\begin{equation}\label{eq:density_matrix}
    \begin{split}
        P_k^{GG'}&=\sum_\nu f_{k\nu}c_{k\nu G}c^*_{k\nu G'},\\
        \frac{1}{N_k}\sum_k\mathrm{Tr}\,P_k&=N_c.
    \end{split}
\end{equation}
Occupations are determined globally across bands and momenta. We do not constrain each momentum to have nine occupied bands. The density coefficients per cell are
\begin{equation}\label{eq:rho_d}
    \rho_d = \frac{1}{N_k}\sum_k \sum_{G-G'=d} P^{GG'}_k, \qquad \rho_0 = N_c.
\end{equation}
In the following equations reciprocal indices denote physical scaled reciprocal vectors. The direct field is
\begin{equation}\label{eq:hartree_field}
    V^H_{GG'} = \begin{cases} \dfrac{2\pi\rho_{G-G'}}{A_c r_s |G-G'|}, & G \neq G' \\ 0, & G = G' \end{cases}
\end{equation}
The exchange field is
\begin{equation}\label{eq:exchange_field}
    \begin{split}
        V^X_{k,GG'}={}&-\frac{2\pi}{A_c r_s N_k}\\
        &\times\sum_{k'}\sum_{H-H'=G-G'}\frac{P^{HH'}_{k'}}{|k-k'+G-H|}.
    \end{split}
\end{equation}
The singular kernel in Eq.~\eqref{eq:exchange_field} is understood as a reciprocal-cell average, described below. The self-consistent one-body matrix is
\begin{equation}\label{eq:fock_matrix}
    \begin{split}
        F_{k,GG'}={}&\frac{|k+G|^2}{2r_s^2}\delta_{GG'}+U_{GG'}\\
        &+V^H_{GG'}+V^X_{k,GG'}.
    \end{split}
\end{equation}
Direct-only Hartree is obtained by setting $V^X = 0$ throughout. For either approximation the energy per electron is
\begin{equation}\label{eq:energy_per_electron}
    \begin{split}
        e={}&\frac{1}{N_c N_k}\sum_k\mathrm{Tr}\left[(T_k+U)P_k\right]\\
        &+\frac{1}{2N_c N_k}\sum_k\mathrm{Tr}\left[(V^H+V^X_k)P_k\right].
    \end{split}
\end{equation}
The factor 1/2 applies to the interaction fields, not to the external potential.

At nonzero temperature we minimize the mean-field free energy per electron $f = e - TS$, where
\begin{equation}\label{eq:entropy}
    \begin{split}
        S=-\frac{k_B}{N_c N_k}\sum_{k\nu}\bigl\{&f_{k\nu}\ln f_{k\nu}\\
        &+(1-f_{k\nu})\ln(1-f_{k\nu})\bigr\}.
    \end{split}
\end{equation}
Diagonalizing $F_k$ gives Fermi--Dirac occupations with one chemical potential fixed by the total particle number. At zero temperature the globally lowest $N_c N_k$ states are filled. Fractional occupations are genuine finite-temperature occupations; the main diagram is labeled $k_B T = 0.005 E_F$.

\subsection{Why clean Hartree remains uniform}\label{sec:clean_hartree}

In the continuum clean problem, the direct energy relative to the neutral uniform state is a sum of nonnegative Fourier contributions proportional to $|\rho_d|^2/|d|$. The filled free-particle states minimize the kinetic energy at fixed particle number. These two bounds are attained simultaneously by the uniform gas, which is therefore a global minimum of direct-only Hartree theory. A clean crystal found in this particular Hartree functional would signal either a constraint or a numerical problem. The argument does not apply to HF because its exchange energy is negative and depends on the off-diagonal density matrix.

For reference, the analytic uniform-gas HF energies are
\begin{equation}\label{eq:uniform_gas_energy}
    \begin{split}
        e^{\mathrm{pol}}_{\mathrm{FG}}&=\frac{1}{r_s^2}-\frac{8}{3\pi r_s},\\
        e^{\mathrm{unpol}}_{\mathrm{FG}}&=\frac{1}{2r_s^2}-\frac{4\sqrt2}{3\pi r_s}.
    \end{split}
\end{equation}
These formulas serve as normalization checks; a uniform stationary solution need not be the unrestricted HF minimum~\cite{bernu2011hartree}.

\section{Quantum numerical implementation and checks}\label{sec:quantum_numerics}

The principal disorder scan uses $M=3$ and 91 plane waves with integer indices satisfying $i^2+j^2+ij \leq 25$. We evaluate exchange using a padded linear Fourier convolution on the combined $k+G$ mesh. Every reciprocal quadrature cell uses an eight-by-eight Gaussian average of the Coulomb kernel; the central singular cell is integrated separately in polar coordinates. This prescription is used consistently in energies and fields.

Self-consistency uses damped iterations and residual extrapolation, followed when necessary by Newton--Krylov refinement. The acceptance criterion is
\begin{equation}\label{eq:scf_residual}
    R = \frac{\|P_{\mathrm{FD}}[F(P)] - P\|_F}{\sqrt{N_c N_k}} < 10^{-6}.
\end{equation}
Each HF point is attempted from a crystal or continuation seed and a noninteracting disordered seed. Difficult points additionally use saved states, reverse continuation in disorder, and variational orbital refinement. We select the lower free energy among converged candidates. The crystal seed is the projector of $T_k+U$ with $-0.05H^*$ added to the elements with $G-G'\in Q_6$ (Sec.~\ref{sec:hf_crossover}). The crystal seed is removed from the Hamiltonian during self-consistency. This procedure samples a limited set of basins and does not prove a global minimum. A root of the self-consistency equations is not, by itself, a stability proof.

For orbital refinement, occupied and virtual reference columns $C_o$ and $C_v$ parameterize a fixed-rank subspace through
\begin{equation}\label{eq:orbital_parametrization}
    Z = C_o + C_v X, \qquad P = Z(1+X^\dagger X)^{-1} Z^\dagger.
\end{equation}
The energy gradient with respect to complex orbital rotations is
\begin{equation}\label{eq:orbital_gradient}
    G_X = \frac{2}{N_c N_k} C_v^\dagger (1-P) F Z (1+X^\dagger X)^{-1}.
\end{equation}
Real and imaginary parts are optimized with L-BFGS. This step finds a starting point in the effectively integer-occupation cases; the final residual is always evaluated against the finite-temperature projector with global occupations. The orbital gradient is independently checked by finite differences.

Independent checks include comparison of the accelerated exchange field with a direct reciprocal-index implementation (maximum difference $2.3 \times 10^{-16}$), and a directional derivative of the finite-temperature free energy (absolute discrepancy $4.8 \times 10^{-11}$ in the test case). The noninteracting limit reproduces the direct one-body diagonalization and prescribed particle number. These checks validate the interaction fields and variational derivatives.

The clean audit also used substantially denser primitive-cell meshes. For a polarized triangular state at $r_s = 4$, the energy changes from approximately $-0.152628$ to $-0.152716$ effective Hartree per electron on increasing the mesh from $24^2$ to $48^2$ with 19 plane waves. A two-point $M^{-2}$ extrapolation gives approximately $-0.152745$, consistent at graphical precision with the published clean result~\cite{bernu2011hartree}. At $r_s = 10$, refining from 19 to 37 plane waves gives a density maximum-to-minimum ratio about 14, comparable to the estimate about 13 reported previously~\cite{trail2003unrestricted}. The two density-contrast estimates agree at the precision of that comparison. These clean commensurate checks do not reproduce the full incommensurate phase diagram. In particular, a uniform selected state at $r_s = 2$ in the small-cell search does not contradict the existence of lower-energy incommensurate states outside the states found by this search. Nor does it conflict with the crystals found for $r_s\geq1.44$~\cite{trail2003unrestricted}, which are unpolarized below $r_s\simeq2.6$~\cite{bernu2011hartree} and hence excluded here. At $r_s=2$, the same $48^2$ primitive-cell audit also relaxes a crystal seed to an essentially uniform density.

At clean $r_s = 10$, clean $r_s = 3$, and disordered $r_s = 3$, $W/E_F = 1$, the selected structural diagnostics remain unchanged to the displayed precision on reducing $k_B T/E_F$ from 0.005 to 0.001 and zero. The contrasts at these three points are approximately 0.59048, 0.06154, and 0.15476, respectively. This establishes temperature insensitivity at the tested points and resolution, not a melting temperature.

At $r_s = 3$, $W/E_F = 1$, increasing the orbital basis from 91 to 127 plane waves changes $f_6$ from 0.6052 to 0.5951 and $C$ from 0.1548 to 0.1570. The selected energy decreases by approximately $7.7\times 10^{-4}$ effective Hartree per electron. The structural label is unchanged, although precision energy claims are not justified by the coarse basis. Increasing the Bloch mesh from $3^2$ to $6^2$ at 91 plane waves gives $f_6 = 0.6055$ and $C = 0.1549$, with an energy change of approximately $6.7\times 10^{-4}$ effective Hartree per electron. Both refinements preserve the structural label at the tested point; the supercell size is fixed throughout this scan.

\begin{figure*}[t]
    \centering
    \includegraphics[width=1.0\textwidth]{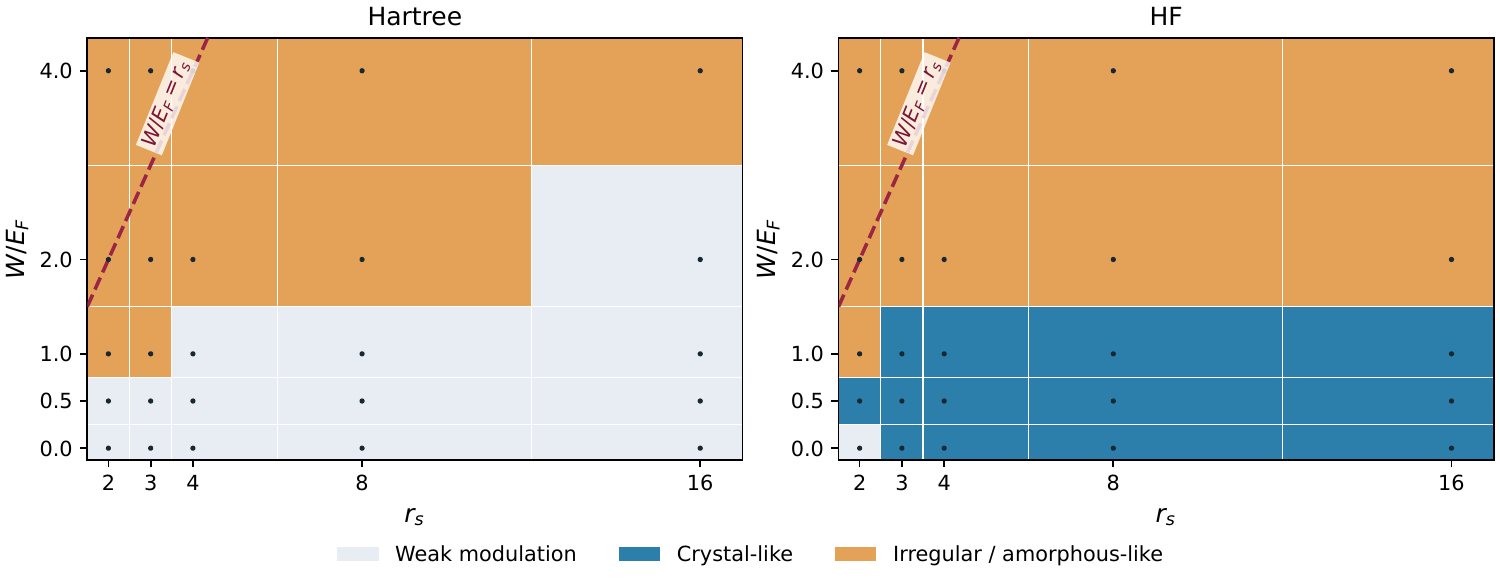}
    \caption{Single-realization structural crossover diagram using the thresholds stated in the text. The potential uses seed 1701, without averaging. Each colored cell represents one computed parameter value; cell edges are plotting boundaries, not fitted critical lines. Dots identify the sampled parameters. The dashed line is $W/E_F=r_s$, corresponding to $W=2H^*/r_s$, twice the Wigner--Seitz Coulomb scale; it is a reference line, not a fitted boundary. Hartree and HF use identical disorder and discretization.}
    \label{fig:phase_diagram}
\end{figure*}

\begin{figure*}[t]
    \centering
    \includegraphics[width=.85\textwidth]{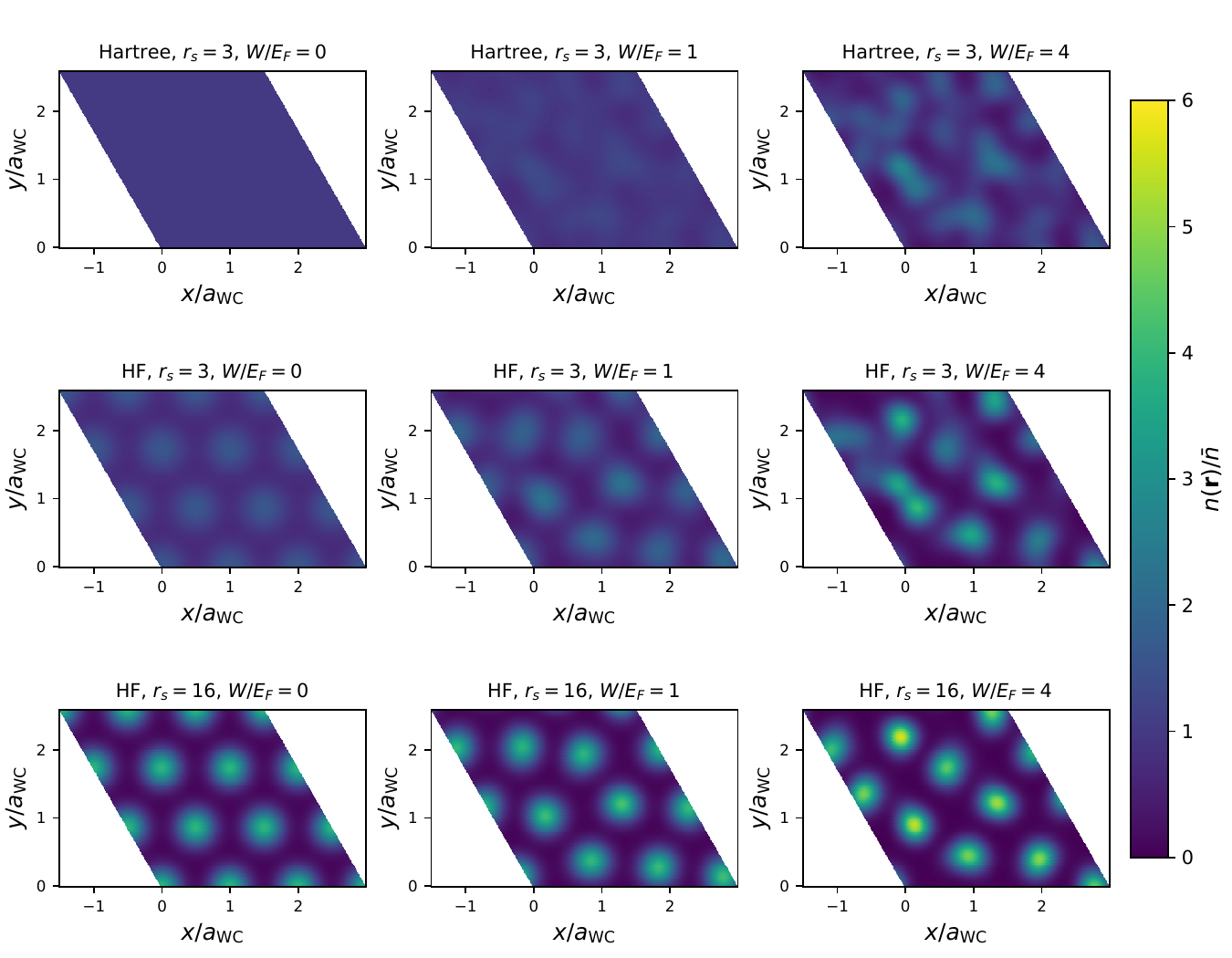}
    \caption{Calculated relative density at the indicated $r_s$ and disorder amplitudes. Coordinates are physical Cartesian distances in units of the pristine lattice constant $a_{\rm WC}$; each rhombus is one supercell. The common color scale permits comparison of Hartree and HF. All panels use the same random-potential realization.}
    \label{fig:density_snapshots}
\end{figure*}

\section{HF structural crossover and diagnostic criteria}\label{sec:hf_crossover}

The density contrast and six-vector Bragg weight are
\begin{equation}\label{eq:density_contrast}
    C = \left\langle (n/\bar n - 1)^2 \right\rangle = \sum_{d\neq 0} |\rho_d/N_c|^2,
\end{equation}
\begin{equation}\label{eq:bragg_fraction}
    B_6 = \sum_{d\in Q_6} |\rho_d/N_c|^2, \qquad f_6 = B_6/C.
\end{equation}
The ratio $f_6$ is defined for $C>0$; a uniform density is assigned the weak-modulation label without evaluating this ratio.

Here $Q_6$ consists of integer supercell vectors $(3,\allowbreak 0),\allowbreak (0,\allowbreak 3),\allowbreak (3,\allowbreak -3)$ and their negatives. If $C < 0.01$ we label the state weakly modulated. Otherwise $f_6 \geq 0.5$ defines crystal-like modulation, and $f_6 < 0.5$ defines irregular or amorphous-like modulation. The weak-modulation cutoff corresponds to a root-mean-square relative density variation below 10\%.

The points in Fig.~\ref{fig:phase_diagram} show a crossover from predominantly crystalline modulation to an increasingly irregular, amorphous-like density. At $r_s = 3$, $f_6$ decreases from approximately 1 in the clean crystal to 0.61 at $W/E_F = 1$ and 0.37 at $W/E_F = 2$. The chosen $f_6 = 0.5$ criterion therefore crosses between these sampled amplitudes, as it also does at $r_s = 4,\allowbreak 8,\allowbreak $ and 16. Changing the criterion to 0.4 moves the latter three brackets to $2 < W/E_F < 4$; a criterion of 0.6 retains the original bracket. At $r_s = 2$, $W/E_F = 0.5$, the selected state has $C \simeq 0.0232$ and $f_6 \simeq 0.517$, close to the convention-dependent boundary. This near-threshold label does not establish disorder-induced thermodynamic crystallization. Crystalline correlations survive weak random distortion, whereas their loss makes the amorphous character more pronounced at stronger disorder. Increasing density contrast can coexist with decreasing coherence, so pronounced modulation alone is insufficient to identify a crystal.

Hartree is uniform at zero disorder and develops inhomogeneity under the external potential. The distinction is visible in Fig.~\ref{fig:density_snapshots}: the HF clean state already has a triangular modulation, whereas the Hartree density modulation is induced by disorder. Labeling a Hartree density profile amorphous-like is a description of its Fourier content, not a claim that a Hartree Wigner crystal has melted.

\begin{figure}[t]
    \centering
    \includegraphics[width=.96\columnwidth]{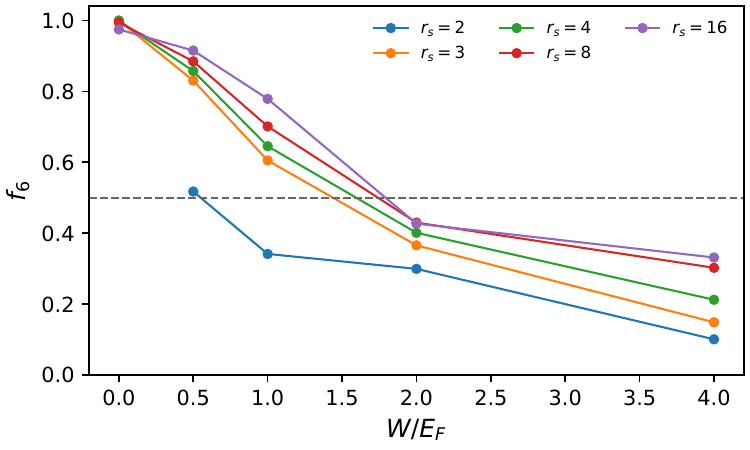}
    \caption{Continuous HF Bragg fractions for the selected states. Points with $C<0.01$ are omitted. Lines join sampled values as a visual guide. The horizontal dashed line is the chosen $f_6=0.5$ threshold; changing this convention changes the apparent structural boundary.}
    \label{fig:bragg_fraction_plot}
\end{figure}

The supercell spans only three pristine lattice spacings along each primitive direction. Section~\ref{sec:coherence} therefore evaluates a density-coherence diagnostic within half that range and distinguishes it from an asymptotic correlation length. The Bragg fraction (Fig.~\ref{fig:bragg_fraction_plot}) and the first-shell coherence measure both quantify the loss of the pristine density-wave pattern. They are orientation selective: a rotated or reconstructed crystal can also lose weight at the fixed reference vectors. The quantum map describes the computed density within one repeated cell; it does not establish the thermodynamic decay of correlations in independently disordered cells.

\section{Classical electrons in an open square}\label{sec:classical_model}

\subsection{Hamiltonian, background, and disorder}\label{sec:classical_hamiltonian}

The classical study treats point electrons with Coulomb interactions and open electrostatic boundaries. Its independently specified Hamiltonian and particle dynamics isolate the competition of Coulomb elasticity, quenched pinning, and thermal motion. Together with the quantum calculation, it tests how the same disordering mechanism acts in two microscopic descriptions of an electron solid. Classical electron crystallization was observed on liquid helium~\cite{grimes1979evidence}, and extensive simulations establish that resolving its bulk melting mechanism requires substantially larger systems and finite-size analysis~\cite{mazars2015melting}. Simulations with random pins also show that structural disordering and thermal effects can compete, including reentrant ordering in some parameter regimes~\cite{reichhardt2023melting}.

Choose a reference density $n_0 = 1/(\pi a_0^2)$, Coulomb energy $E_0 = e^2/(4\pi\epsilon a_0)$, and time $t_0 = (ma_0^2/E_0)^{1/2}$. Coordinates, energies, and times below are in these units. The control parameters are $q = n_e/n_0$, $c = \bar n_i/n_0$, and $\theta = k_B T/E_0$. With $N = 288$ electrons, the square has side $L = (\pi N/q)^{1/2}$. The local Wigner--Seitz radius is $a_0/\sqrt q$, the triangular spacing is $a = (2\pi/\sqrt 3)^{1/2}/\sqrt q$, and the local Coulomb scale is $E_C = E_0\sqrt q$. Thus $\Gamma = \sqrt q/\theta$. These classical parameters contain no Planck constant; a classical density sweep does not describe quantum melting at fixed $r_s$.

The dimensionless configurational energy is
\begin{equation}\label{eq:U_classical}
    \begin{split}
        U={}&\sum_{i<j}\frac{1}{r_{ij}}-\frac{q}{\pi}\sum_i I_L(x_i,y_i)\\
        &+\sum_{i\alpha}v_p(|r_i-R_\alpha|).
    \end{split}
\end{equation}
\begin{equation}\label{eq:background_integral}
    I_L(x,y) = \int_0^L dX \int_0^L dY\, [(x-X)^2+(y-Y)^2]^{-1/2}.
\end{equation}
The positive uniform background carries charge $+Ne$. Its position-independent self-energy is omitted. Reflecting walls constrain the electrons to the square. There are no repeated cells or image charges. Every electron pair is evaluated with the untruncated $1/r$ interaction, without short-distance softening. Conventional periodic Ewald summation would define a different boundary-value problem; for this finite open system, the direct all-pair sum is exact.

The background integral is evaluated analytically. Define
\begin{equation}\label{eq:asinh_kernel}
    A(u,v) = u\,\mathrm{asinh}(v/|u|) + v\,\mathrm{asinh}(u/|v|),
\end{equation}
with continuous limiting values on the axes, $u_0=-x$, $u_1=L-x$, $v_0=-y$, $v_1=L-y$. Then
\begin{equation}\label{eq:background_analytic}
    \begin{split}
        I_L={}&A(u_1,v_1)-A(u_0,v_1)\\
        &-A(u_1,v_0)+A(u_0,v_0).
    \end{split}
\end{equation}
For example, writing $J(d;y) = \mathrm{asinh}(y/d) + \mathrm{asinh}[(L-y)/d]$, the background force is $F_x^{\mathrm{bg}} = (q/\pi)[J(x;y) - J(L-x;y)]$, with $F_y^{\mathrm{bg}}$ obtained by exchanging $x$ and $y$. This force attracts electrons toward the interior and balances the macroscopic repulsion.

The short-range attractive impurity wells are
\begin{equation}\label{eq:pin_potential}
    \begin{split}
        v_p(r)&=-D(1-r^2/R_p^2)^2\,\Theta(R_p-r),\\
        D&=0.1,\qquad R_p=0.4(2\pi/\sqrt3)^{1/2}.
    \end{split}
\end{equation}
Both potential and force vanish continuously at the cutoff. Inside a well,
\begin{equation}\label{eq:pin_force}
    F^p_{i\alpha} = -\frac{4D}{R_p^2}(1-r_{i\alpha}^2/R_p^2)(r_i - R_\alpha).
\end{equation}
The electron-electron force is $\sum_{j\neq i}(r_i-r_j)/r_{ij}^3$. The well depth and radius remain fixed in physical units when either density is varied. In particular, $D/E_C = 0.1/\sqrt q$ and $R_p/a = 0.4\sqrt q$ vary with electron density.

We generate one marked Poisson impurity field with seed 1701 over the largest square, at nominal density $4n_0$. Each point has an independent uniform mark in $[0,\allowbreak 1]$. At concentration $c$, points with marks below $c/4$ are retained; at electron density $q$, the field is restricted to the corresponding concentric square. This gives nested disorder in the same physical coordinate system. The actual number $N_i$ fluctuates around $\bar n_i L^2$. There is no disorder averaging.

\subsection{Zero-temperature relaxation and finite-temperature dynamics}\label{sec:classical_dynamics}

The density grid is $q = 0.5,\allowbreak 0.75,\allowbreak 1,\allowbreak 1.5,\allowbreak 2$ and $c = 0,\allowbreak 1/64,\allowbreak 1/32,\allowbreak 1/16,\allowbreak 1/8,\allowbreak 1/4,\allowbreak 1/2,\allowbreak 1,\allowbreak 2,\allowbreak 4$. At $T=0$ we use the FIRE damped-inertial relaxation method~\cite{bitzek2006structural}. The initial clean configuration is an exact triangular lattice with 16 columns and 18 rows. The open square distorts its outer rows after relaxation. For each disordered point we compare relaxation from this lattice and, where available, continuation from a lower impurity concentration, selecting the lower energy among converged states. All 50 points satisfy $\max_{i\mu}|F_{i\mu}| < 10^{-7}E_0/a_0$. This is a specified local-minimum search, not proof of a global classical ground state. Other starting configurations reach nearly degenerate minima (relative energy differences below $10^{-3}$) that change at most one of the 50 structural labels; near the structural crossover, however, their order parameters and coherence lengths can differ appreciably, so single-state values quoted there are representative rather than precise.

At finite temperature, unit-mass electrons obey
\begin{equation}\label{eq:langevin}
    \begin{split}
        \ddot r_i&=F_i-\gamma\dot r_i+\sqrt{2\gamma\theta}\,\eta_i(t),\\
        \gamma&=0.5,
    \end{split}
\end{equation}
where independent noise components satisfy $\langle \eta_{i\mu}(t)\eta_{j\nu}(t')\rangle = \delta_{ij}\delta_{\mu\nu}\delta(t-t')$. We use the BAOAB Langevin splitting~\cite{leimkuhler2013rational}, with step $h=0.04$:
\begin{equation}\label{eq:baoab}
    \begin{split}
        v&\leftarrow v+(h/2)F,\quad r\leftarrow r+(h/2)v,\\
        v&\leftarrow e^{-\gamma h}v+[\theta(1-e^{-2\gamma h})]^{1/2}\zeta,\\
        r&\leftarrow r+(h/2)v,\quad v\leftarrow v+(h/2)F(r).
    \end{split}
\end{equation}
Here $\zeta$ is a fresh standard normal vector. Specular wall reflections are applied during each drift and after each FIRE position update; Coulomb separations are never wrapped. Each concentration is heated independently from its own relaxed state at $q=1$. The 12 temperatures are $\theta = 0.001,\allowbreak 0.002,\allowbreak 0.0035,\allowbreak 0.005,\allowbreak 0.007,\allowbreak 0.010,\allowbreak 0.014,\allowbreak 0.020,\allowbreak 0.030,\allowbreak 0.050,\allowbreak 0.080,\allowbreak $ and $0.120$. Each point has 300 $t_0$ for equilibration and 600 $t_0$ for measurement, giving 120 temperature points. Configurations are saved every $t_0$. Thermal time averages at fixed impurities are distinct from disorder averages.

Independent validation compares the force with a directional energy derivative (absolute error $1.4\times 10^{-9}$) and the analytic background integral with numerical quadrature (error $1.5\times 10^{-14}$). Undamped dynamics in a 36-particle test has a total-energy range per particle of $1.3\times 10^{-6}E_0$ at $h=0.04$, reduced to $4.1\times 10^{-7}E_0$ at $h=0.02$. A separate audit recomputes forces and energies at all 50 relaxed states using the same force implementation. The four configurations displayed below have positive minimum Hessian eigenvalues, approximately 0.0032, 0.0046, 0.0107, and 0.1324 in units $E_0/a_0^2$, confirming their local mechanical stability while leaving the limited basin search explicit.

\section{Classical structural crossover and density correlations}\label{sec:classical_crossover}

\subsection{Order measures and open-window corrections}\label{sec:order_measures}

To reduce boundary contributions, structural observables use electrons at least $2a$ from every wall. Neighbors and coordination numbers $z_i$ come from a Delaunay triangulation of all electrons. Define
\begin{equation}\label{eq:psi6}
    \begin{split}
        \psi_6(i)&=\frac{1}{z_i}\sum_{j\in N_i}e^{6i\phi_{ij}},\\
        \Psi_6&=\left|\frac{1}{N_I}\sum_{i\in I}\psi_6(i)\right|,\qquad
        P_6=\frac{1}{N_I}\sum_{i\in I}\delta_{z_i,6}.
    \end{split}
\end{equation}
The interior set $I$ has side $\ell = L-4a$ and population $N_I$. Global $\Psi_6$ is insensitive to a rigid lattice rotation. Together with $P_6$, it distinguishes orientational coherence from the local coordination count. We call a sampled state crystal-like if $\Psi_6 \geq 0.5$ and $P_6 \geq 0.85$. These numerical thresholds specify the map; they do not identify a thermodynamic singularity or resolve a hexatic phase.

For density correlations we use a translation correction for the open observation window. For an interior pair with displacement $(\Delta x,\allowbreak \Delta y)$, set $w_{ij} = [(1-|\Delta x|/\ell)(1-|\Delta y|/\ell)]^{-1}$. For annulus $b$ of area $A_b$, the estimator is
\begin{equation}\label{eq:pair_correlation}
    g_b = \frac{2\ell^2}{N_I(N_I-1)A_b}\sum_{i<j\in I} w_{ij}\mathbf{1}_b(r_{ij}).
\end{equation}
It tends to unity for uniformly independent points. We also compute $g_6(r) = \langle \psi_6(i)\psi_6(j)^*\rangle_r$ with the same pair weights, reporting its real part. Empty distance bins are left undefined. Only separations through $5a$, less than half the interior side, are shown.

The finite-window scattering intensity subtracts the uniform square form factor:
\begin{equation}\label{eq:Sw}
    S_w(k) = \frac{1}{N_I}\left|\sum_{i\in I} e^{-ik\cdot(r_i-r_c)} - N_I f_\ell(k)\right|^2,
\end{equation}
\begin{equation}\label{eq:window_form_factor}
    \begin{split}
        f_\ell(k)&=\mathrm{sinc}(k_x\ell/2)\,\mathrm{sinc}(k_y\ell/2),\\
        \mathrm{sinc}(x)&=\sin x/x.
    \end{split}
\end{equation}
This removes the deterministic forward peak of the observation square. It is an instantaneous, window-corrected density intensity, including the static pinned structure --- not a connected thermal fluctuation spectrum. No periodic continuation or synthetic density pattern is used.

\begin{figure*}[t]
    \centering
    \includegraphics[width=1.0\textwidth]{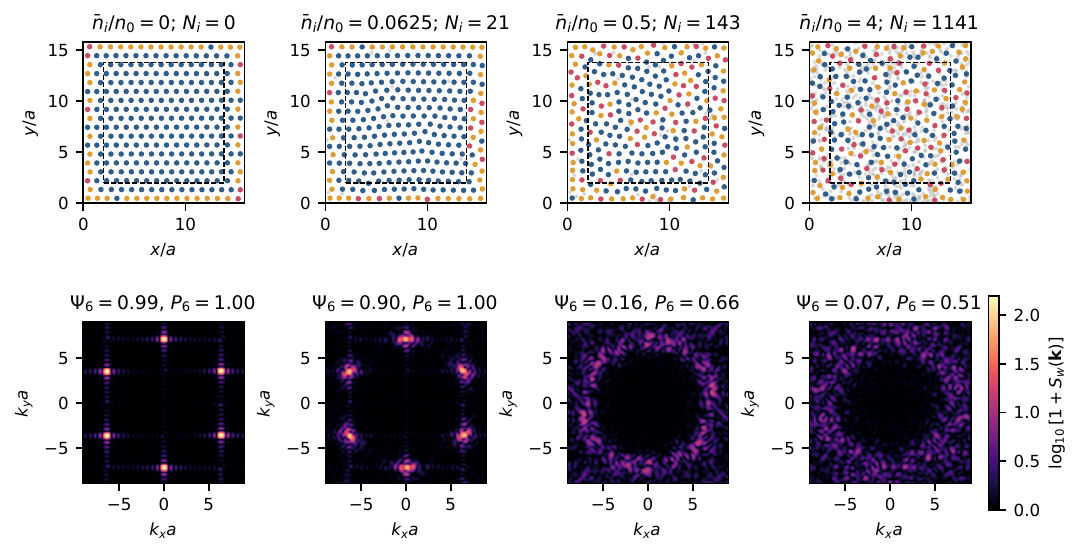}
    \caption{Classical $T=0$ configurations at $n_e=n_0$ and increasing impurity density. Blue particles have six Delaunay neighbors, orange fewer, and red more; faint gray points are impurities. Dashed squares show the interior observation region. The lower row is the calculated window-corrected structure factor, Eq.~\eqref{eq:Sw}, on a common logarithmic color scale. The sharp sixfold peaks broaden and become diffuse as the relaxed structure becomes irregular. All panels use one nested impurity field.}
    \label{fig:classical_snapshots_structure}
\end{figure*}

\begin{figure*}[t]
    \centering
    \includegraphics[width=.96\textwidth]{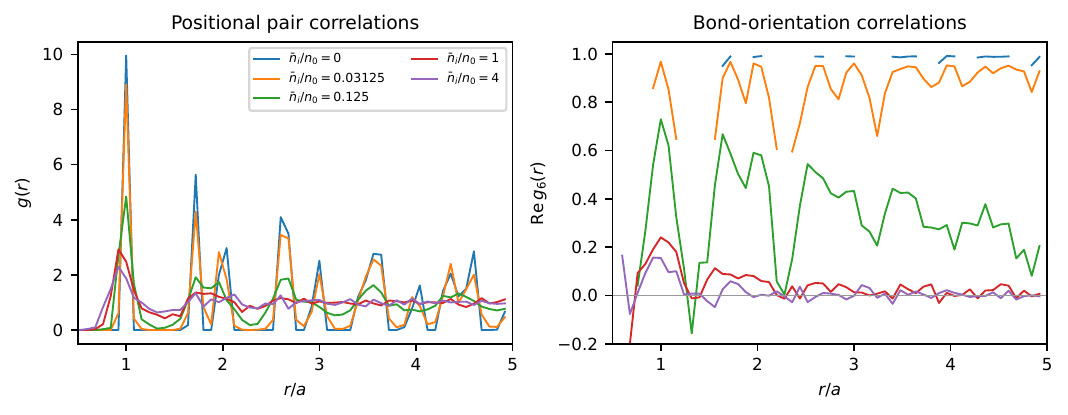}
    \caption{Positional and bond-orientation correlations at $T=0$ and $n_e=n_0$. Increasing the impurity concentration broadens the sharp coordination shells and suppresses orientational correlations at larger separation. These are spatial pair averages within one quenched realization; the sampling window and its edge correction are specified in the text.}
    \label{fig:classical_correlations}
\end{figure*}

\subsection{Disorder evolution and dependence on electron density}\label{sec:classical_density_dependence}

The clean square has $P_6=1$ and $\Psi_6=0.992$ in the interior. Figure~\ref{fig:classical_snapshots_structure} follows the transformation at fixed electron density. At $c=1/16$, the selected configuration still has $P_6=1$ but $\Psi_6\simeq 0.90$: pinning distorts bond angles before creating interior coordination defects. Earlier simulations likewise report a defect-free crystal at weak pinning~\cite{reichhardt2022nonlinear}, dislocations followed by disclinations as disorder increases~\cite{cha1994orientational,cha1994topological}, and use $P_6$ to track this disordering~\cite{reichhardt2001moving,reichhardt2023melting}. At $c=1/2$, $P_6\simeq 0.66$ and $\Psi_6\simeq 0.16$, and at $c=4$ these become approximately 0.51 and 0.07. The actual impurity counts in these four squares are 0, 21, 143, and 1141. The large-disorder states are mechanically stable and amorphous-like over the available observation window. The weakly pinned state is already distorted from the pristine lattice, while stronger pinning adds defects and erodes its remaining coherence.

The structure factors provide a complementary diagnosis. Weak disorder preserves six identifiable peaks, with increased diffuse scattering; at stronger disorder, the intensity spreads into a broad shell, as found for classical Wigner crystals with charged impurities~\cite{huang2026interplay} and imaged in Ref.~\cite{ge2026visualizing}. Correspondingly, $g(r)$ retains a short-distance exclusion region and a broadened nearest-neighbor peak after its distant coordination shells have weakened. The bond-orientation correlation also falls toward zero at larger distances (Fig.~\ref{fig:classical_correlations}). Thus local packing survives the loss of coherent crystal order. Section~\ref{sec:coherence} quantifies this shortening with an operational coherence length, retaining explicit bounds whenever the decay exceeds the observation window.

\begin{figure*}[t]
    \centering
    \includegraphics[width=.96\textwidth]{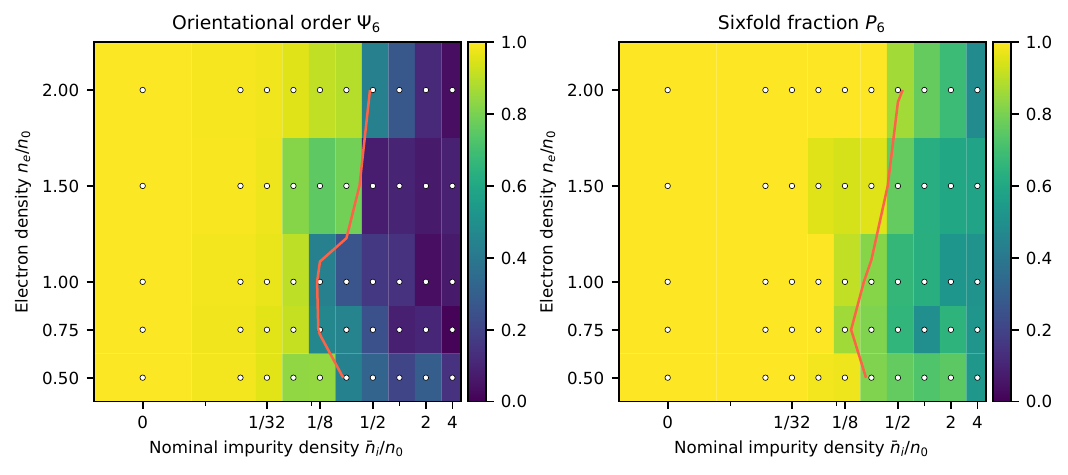}
    \caption{Classical impurity-density--electron-density crossover diagram at $T=0$, with fixed physical well radius and depth. Fifty computed points are shown by dots. Colors give $\Psi_6$ and $P_6$; red contours are interpolated guides at 0.5 and 0.85, respectively. Crystal-like states satisfy both thresholds. The horizontal axis uses a linear scale near zero and logarithmic compression above $1/64$. The nominal impurity density is the intensity of the single marked Poisson process. The contours are finite-sample structural crossovers.}
    \label{fig:classical_density_diagram}
\end{figure*}

Figure~\ref{fig:classical_density_diagram} shows the two-density map. Raising electron density strengthens the Coulomb scale but also changes the pin radius relative to the lattice spacing; increasing impurity density changes the number of available wells. At $q=2$, configurations at $c=1/8$ and $1/4$ retain $P_6\simeq1$, whereas at $q=1$ the corresponding states have already lost substantial coherence. Individual threshold locations fluctuate with the finite impurity pattern and need not be monotone in density. The contours give the structural crossovers for the specified well depth, radius, and open-square geometry.

\section{Classical thermal disordering and cage escape}\label{sec:thermal}

\begin{figure*}[t]
    \centering
    \includegraphics[width=1.0\textwidth]{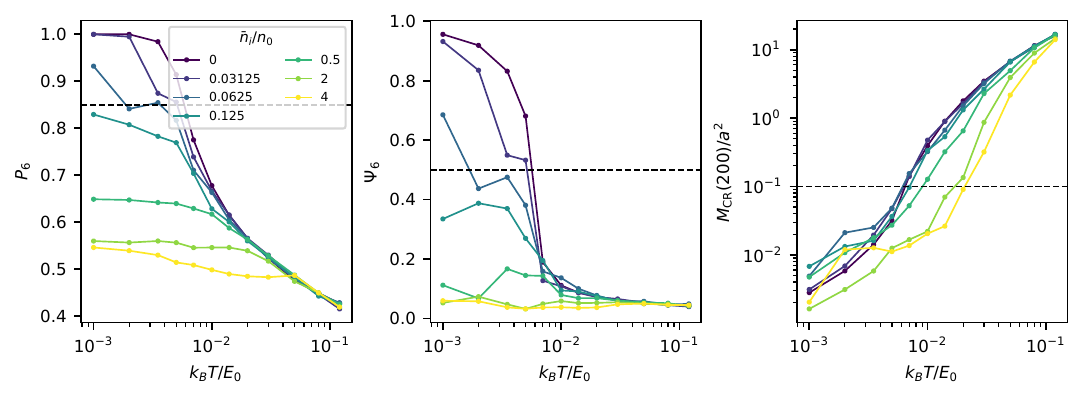}
    \caption{Thermal evolution at $n_e=n_0$: coordination, global bond orientation, and cage-relative displacement. Dashed lines show the stated diagnostic thresholds. Each curve is a heating trajectory for one impurity concentration. Crystalline order can disappear while rearrangements remain slow; the latter are increasingly suppressed at strong pinning.}
    \label{fig:classical_thermal_curves}
\end{figure*}

An amorphous pinned state has no crystalline Bragg order to lose. A single temperature labeled ``melting'' would therefore conflate structural melting with thermally activated rearrangements. We report two operational temperatures: $T_{\mathrm{str}}$, where the crystal-like criterion ceases to hold during heating, and $T_{\mathrm{mob}}$, where relative cage motion crosses a fixed threshold (Fig.~\ref{fig:classical_thermal_curves}). If no sampled temperature is crystal-like, no $T_{\mathrm{str}}$ is assigned.

\begin{figure}[t]
    \centering
    \includegraphics[width=1.0\columnwidth]{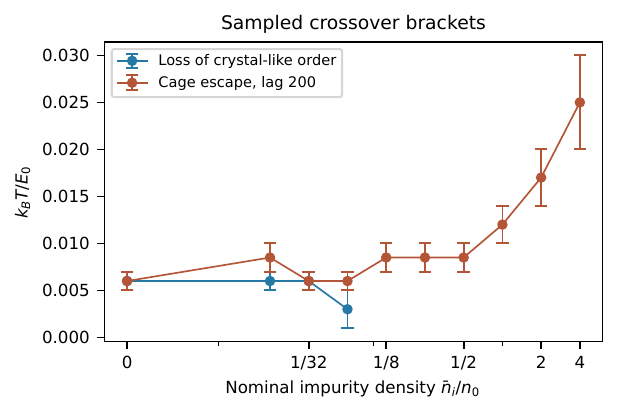}
    \caption{Temperatures for loss of crystal-like order and appreciable cage escape versus impurity concentration. The structural bracket at $c=1/16$ includes the sampling dependence discussed in the text; other bars are adjacent-temperature brackets of the primary protocol. The structural branch terminates when no ordered interval is found. The mobility branch continues through the amorphous pinned regime and depends on observation time. No thermodynamic melting line is assigned there.}
    \label{fig:classical_melting}
\end{figure}

At each time origin $\tau$, let $N_i^\tau$ be the six nearest particles and $\Delta r_i(t;\tau) = r_i(\tau+t)-r_i(\tau)$. The cage-relative mean-square displacement~\cite{mazoyer2010dynamics,illing2017mermin,vivek2017longwavelength}, a dynamic Lindemann measure~\cite{bedanov1985modified}, is
\begin{equation}\label{eq:cage_msd}
    \begin{split}
        M_{\mathrm{CR}}(t)=\biggl\langle\biggl|&\Delta r_i(t;\tau)\\
        &-\frac{1}{6}\sum_{j\in N_i^\tau}\Delta r_j(t;\tau)\biggr|^2\biggr\rangle_{i,\tau}.
    \end{split}
\end{equation}
Particles are selected in the interior at the origin; their subsequent motion is not censored if they leave it. The reference cage stays fixed during each lag. We define $T_{\mathrm{mob}}$ by $M_{\mathrm{CR}}(200t_0)/a^2 = 0.1$, and repeat the analysis at lags 100 and 400 $t_0$. This is a finite-observation-time mobility crossover, not a diffusion coefficient or a thermodynamic glass transition. In fact, there is no glass in the presence of explicit external disorder in the Hamiltonian since the Hamiltonian lacks translational invariance. Symbols in Fig.~\ref{fig:classical_melting} denote bracket midpoints. Bars span sampled crossing intervals, broadened to reflect the structural sampling dependence at $c=1/16$; they are not statistical confidence intervals.

Figure~\ref{fig:classical_thermal_curves} shows that the clean $N=288$ system changes from $P_6\simeq 0.91$, $\Psi_6\simeq 0.68$ at $\theta=0.005$ to $P_6\simeq 0.78$, $\Psi_6\simeq 0.19$ at $\theta=0.007$. Its primary structural and mobility crossovers therefore both lie between these temperatures. The experimental clean classical scale $\Gamma \simeq 137\pm 15$~\cite{grimes1979evidence}, close to Monte Carlo and elastic-theory estimates~\cite{gann1979monte,morf1979temperature}, corresponds to temperatures of this order, but our open finite system and operational thresholds do not reproduce a bulk melting point to that precision.

At $c=1/8$ and higher, no sampled heating point satisfies the crystal-like criterion, so extending a crystalline melting line across this region would be misleading. Nevertheless, the cage-escape scale rises. In the primary protocol it is bracketed by $\theta = 0.010$--$0.014$ at $c=1$, by $0.014$--$0.020$ at $c=2$, and by $0.020$--$0.030$ at $c=4$ (Fig.~\ref{fig:classical_melting}). This increase concerns kinetic unfreezing of an already disordered assembly.

The dilute case $c=1/16$ is particularly sensitive to sampling. The primary heating trajectory loses the chosen structural label between 0.001 and 0.002. Extended runs recover crystal-like order at 0.002; at 0.0035 an extended run gives $\Psi_6 \simeq 0.515$ and $P_6\simeq 0.851$, both close to the diagnostic thresholds; at 0.005 the extended state is clearly disordered. We therefore show the wider 0.001--0.005 sampling envelope for its structural crossover. These trajectories do not resolve a reentrant ordered interval or a precise melting temperature. Such reentrance at weak disorder was predicted by Nelson~\cite{nelson1983reentrant} and found in Ref.~\cite{reichhardt2023melting}.

All 120 kinetic-temperature averages agree with their targets to within 1\%. Halving the time step at $(c,\allowbreak \theta)=(0,\allowbreak 0.007)$ and $(1,\allowbreak 0.014)$ preserves their disordered classification and gives $M_{\mathrm{CR}}(200)/a^2 \simeq 0.125$ and $0.110$, respectively, above the mobility threshold. Doubling equilibration and measurement times at selected points preserves the broad trends but exposes borderline values: for the clean system at 0.007 the displacement is approximately 0.100. At $c=1$, extended sampling gives approximately 0.054 at 0.010 and 0.143 at 0.014. At $c=4$, extended sampling gives approximately 0.079 at 0.020 and 0.311 at 0.030, retaining the primary cage-escape bracket. The longer-lag results also shift the apparent crossover, as expected for activated motion.

A separate clean $N=572$ calculation gives $\Psi_6 \simeq 0.70,\allowbreak 0.49,\allowbreak $ and $0.089$ at $\theta = 0.005,\allowbreak 0.007,\allowbreak $ and $0.010$. At 0.007 its 200-$t_0$ cage displacement is only about 0.068, compared with 0.142 in the primary smaller-system run; at 0.010 it rises to about 0.47. Thus size and sampling can shift the apparent mobility threshold by a temperature interval. The defensible clean scale from these checks is roughly $\theta = 0.005$--$0.010$, not a high-precision number. The plotted main curves refer specifically to the stated $N=288$ heating protocol.

The two tendencies have different origins. Random wells disrupt positional and orientational coherence, reducing the stability range of a recognizable crystal. They also create barriers to particle rearrangement, allowing an already irregular configuration to remain caged to higher temperatures. This distinction gives a classical counterpart to the concern that an apparent solid-like insulating regime can persist without a pristine Wigner lattice~\cite{babbar2026wigner,vu2022thermal}. Classical pinning here involves potential barriers and Coulomb forces; it is not a calculation of quantum Anderson interference.

\begin{figure*}[t]
    \centering
    \includegraphics[width=1.0\textwidth]{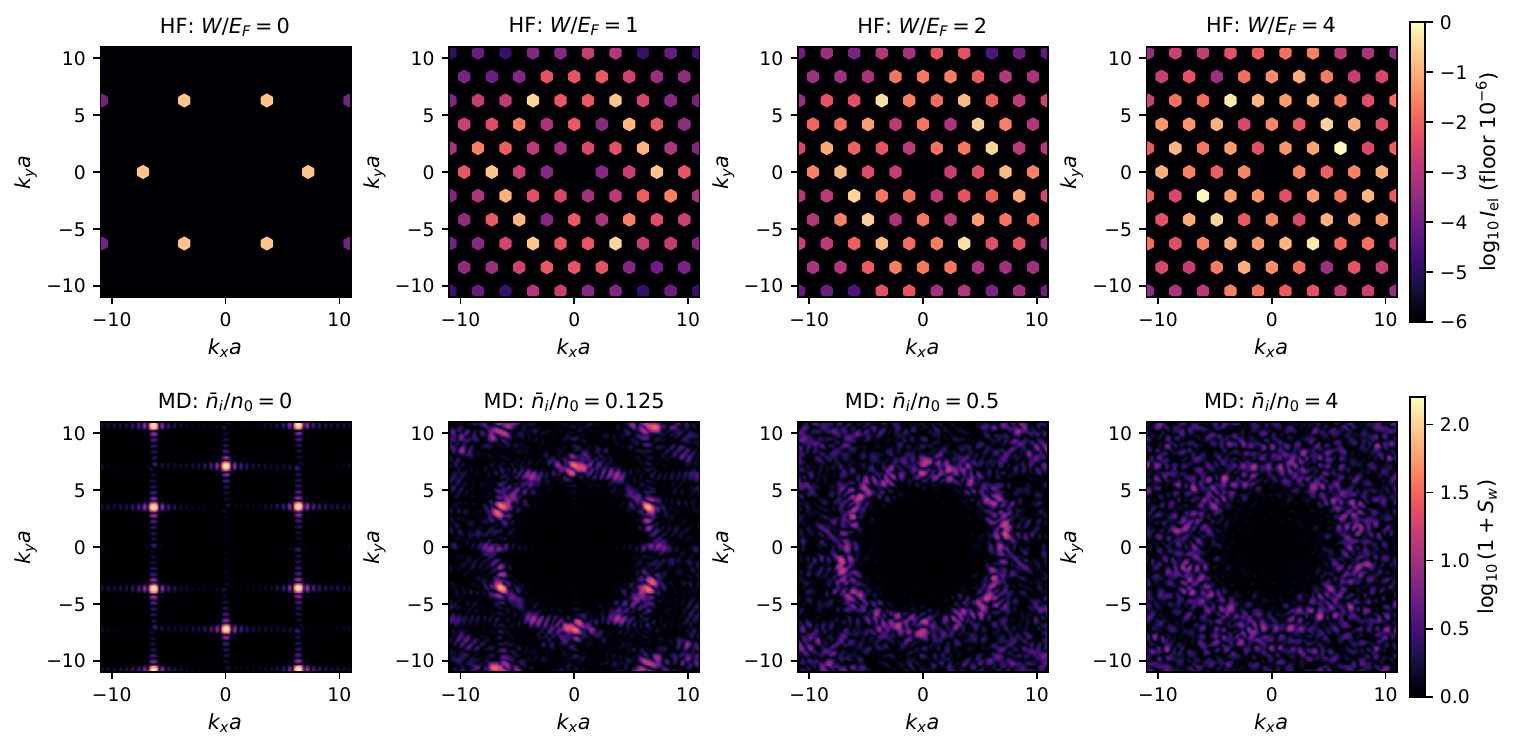}
    \caption{Static density spectra versus disorder in the quantum and classical calculations. Upper row: discrete HF elastic intensity $I_{\mathrm{el}}$ at $r_s=4$, with $W/E_F=0,\allowbreak 1,\allowbreak 2,\allowbreak 4$; only actual reciprocal vectors are plotted, with a logarithmic display floor $10^{-6}$. Lower row: classical window-corrected $S_w$ at $q=1$ and $c=0,\allowbreak 1/8,\allowbreak 1/2,\allowbreak 4$. Each row has a common color scale; the different normalizations are defined in the text. The classical peaks broaden into diffuse scattering, while the periodic quantum spectrum transfers weight between discrete modes. No synthetic linewidth or disorder averaging is introduced.}
    \label{fig:disorder_structure_factors}
\end{figure*}

\section{Structure factors and disorder-dependent coherence}\label{sec:coherence}

\subsection{Quantum and classical density spectra}\label{sec:spectra}

We compute the Fourier spectrum and the same first-shell density-coherence diagnostic from every selected Hartree/HF density matrix and every relaxed classical configuration. Let $a$ denote the pristine triangular spacing at the corresponding density, and $G_1=4\pi/(\sqrt3 a)$ the magnitude of its first reciprocal shell. Distances are reported in units of $a$ separately for each model.

For the quantum calculation, the elastic density intensity at a nonzero supercell reciprocal vector is
\begin{equation}\label{eq:I_elastic}
    I_{\mathrm{el}}(d)=\frac{|\rho_d|^2}{N_c},\qquad d\ne0.
\end{equation}
It measures the static spatial organization of the self-consistent density. This is the coherent contribution built from $|\langle\hat\rho_d\rangle|^2$, with a per-cell normalization. It is not the full equal-time quantum structure factor $\langle\hat\rho_d\hat\rho_{-d}\rangle/N$, which also contains a connected exchange and fluctuation term. The latter is not evaluated here. The classical spectrum is the instantaneous particle intensity $S_w$ of Eq.~\eqref{eq:Sw}, including its particle self term and the uniform-window subtraction. These two spectra address the static density pattern; their absolute intensities are not directly equated.

Figure~\ref{fig:disorder_structure_factors} places the calculated quantum and classical spectra side by side. In HF, disorder transfers weight from the pristine sixfold shell to other supercell reciprocal vectors. Periodic boundary conditions keep these spectral lines discrete; no artificial peak broadening is applied. In MD, increasing pin density broadens the initially sharp peaks into a diffuse shell and then weakens the angular organization of that shell.

\subsection{A common first-shell correlation construction}\label{sec:first_shell}

The oscillations of an ordinary density autocorrelation contain the lattice period as well as the decay of crystalline coherence. To separate these, we shift each of the six pristine Bragg carriers $Q_j$ to zero before transforming its neighboring spectral weight, a band-limited variant of the translational correlation function $\langle e^{iG\cdot(r_i-r_j)}\rangle$~\cite{murray1987experimental,grier1991translational}. Define disjoint patches
\begin{equation}\label{eq:carrier_patch}
    \begin{split}
        \mathcal P_j&=\{k:|k-Q_j|\leq Q_c,\\
        &\qquad |k-Q_j|=\min_l|k-Q_l|\},\qquad Q_c=\alpha G_1,
    \end{split}
\end{equation}
with a fixed tie rule at patch boundaries. The main analysis uses $\alpha=0.5$; $0.4$ and $0.6$ test the bandwidth dependence. For a discrete spectrum, define
\begin{equation}\label{eq:CB_definition}
    A_w(r)C_B(r)=\frac{\sum_j\sum_{k\in\mathcal P_j}I(k)J_0(|k-Q_j|r)}{\sum_j\sum_{k\in\mathcal P_j}I(k)}.
\end{equation}
The sums become uniform quadrature integrals for the classical continuous spectrum. Here $I=I_{\mathrm{el}}$ for HF and $I=S_w$ for MD. The Bessel function is the angular average of the inverse Fourier transform after demodulation; it is not a disorder average. Thus $C_B$ measures the persistence of the first-shell density-wave pattern without its rapid lattice oscillation.

For periodic HF, $A_w=1$. For the classical square observation window, we divide by its angularly averaged overlap fraction,
\begin{equation}\label{eq:window_overlap}
    A_w(r)=1-\frac{4r}{\pi\ell}+\frac{r^2}{\pi\ell^2},\qquad r<\ell.
\end{equation}
This removes the leading geometric loss of pairs with separation. Band filtering and window truncation do not commute exactly, so residual window effects remain; the clean result and the bandwidth checks specify their practical size. We use separations through $5a$, below half the interior side $\ell\simeq11.79a$, and an $81\times81$ quadrature mesh in each carrier patch.

The operational translational coherence length is the first crossing
\begin{equation}\label{eq:xi_B}
    \xi_B=\inf\{r>0:C_B(r)\leq e^{-1}\}.
\end{equation}
We interpolate only between adjacent sampled separations, and never impose an exponential fit. A curve that remains above $e^{-1}$ gives a lower bound: $\xi_B>1.5a$ for the quantum cell and $\xi_B>5a$ for the classical window. A uniform Hartree or HF density has no density-wave coherence length under this definition; its value is undefined, not zero. The numerical threshold for resolving any modulation in this analysis is $C=10^{-8}$, distinct from the display classification $C=0.01$.

The quantum resolution has a particularly simple consequence. At the main bandwidth, the available offsets in each patch have magnitude zero or $G_1/3$. Therefore
\begin{equation}\label{eq:CB_HF}
    C_B^{\mathrm{HF}}(r)=p_0+(1-p_0)J_0(G_1r/3),
\end{equation}
where $p_0$ is the fraction of the selected patch intensity exactly at the six pristine carriers. The HF length consequently parameterizes an intra-cell redistribution of density-wave weight. It is not an independent bulk decay-length fit. In particular, a periodically repeated density cannot supply the asymptotic correlation function of an infinite random system. This concrete resolution limit is retained in the figure labels and in the interpretation of the quantum values.

\begin{figure*}[t]
    \centering
    \includegraphics[width=1.0\textwidth]{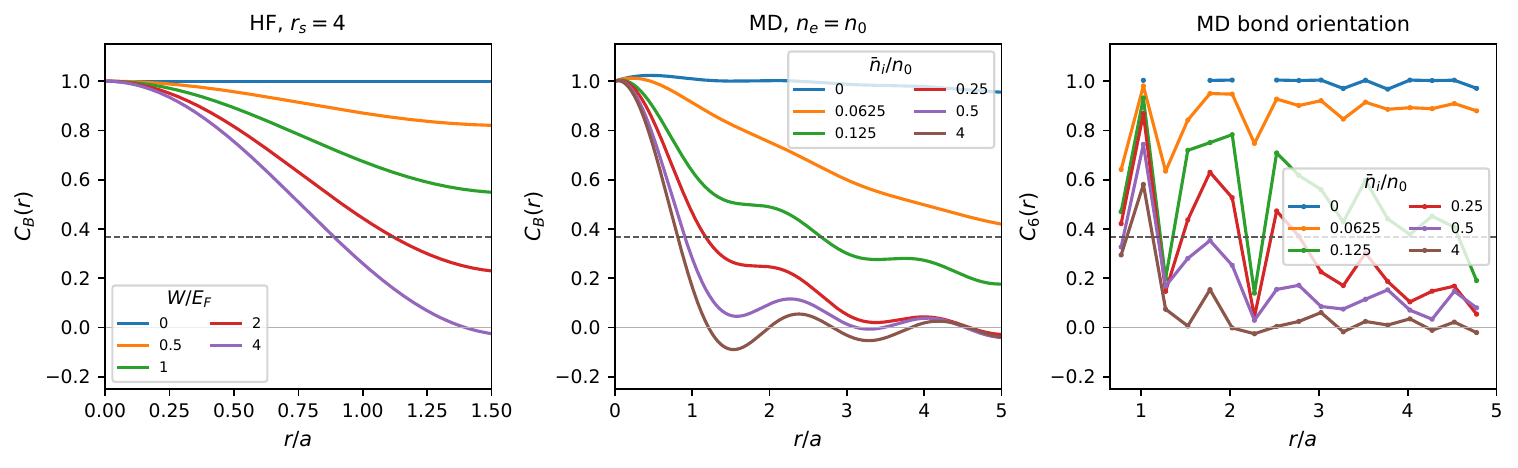}
    \caption{Calculated correlation functions at increasing disorder. Left: HF demodulated first-shell correlation $C_B$ at $r_s=4$. Center: the corresponding classical quantity at $q=1$, including the square-overlap correction. Right: the normalized classical bond-orientation function $C_6$. The dashed horizontal line is $e^{-1}$. Fourier demodulation removes the pristine carrier oscillation; the retained patch bandwidth sets the shortest resolved scale. The classical bond-orientation curves retain coordination-shell oscillations, so their first crossing is not used as a robust decay length.}
    \label{fig:translational_orientational_correlations}
\end{figure*}

\begin{figure*}[t]
    \centering
    \includegraphics[width=.92\textwidth]{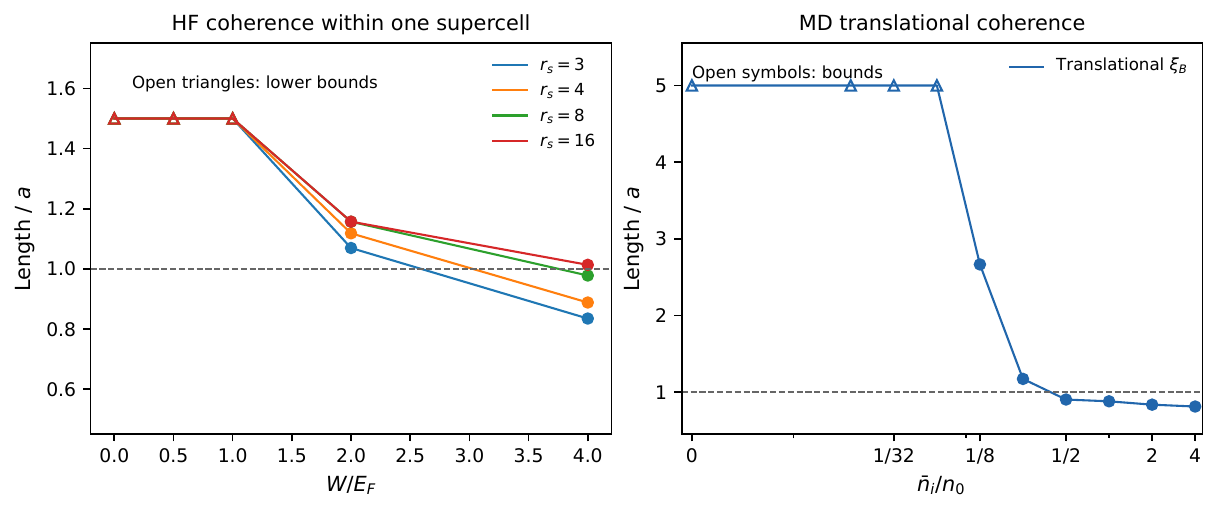}
    \caption{Translational coherence versus disorder for HF and MD, using the same carrier-patch convention $Q_c/G_1=0.5$. Open upward triangles give lower bounds from the accessible range, not measured plateaus. Filled circles are first $e^{-1}$ crossings. The dashed line is one pristine lattice spacing. The quantum values are intra-cell diagnostics within a cell of side three spacings; the classical values use an interior square of side approximately $11.79a$. Lines only connect computed values. Both calculations reach microscopic coherence at strong pinning.}
    \label{fig:correlation_lengths_vs_disorder}
\end{figure*}

\subsection{Calculated correlation lengths and density dependence}\label{sec:correlation_lengths}

Figure~\ref{fig:translational_orientational_correlations} shows the actual correlation curves, and Fig.~\ref{fig:correlation_lengths_vs_disorder} summarizes their crossing lengths. For HF at $r_s=4$, $\xi_B$ exceeds the accessible $1.5a$ at $W/E_F=0,\allowbreak 0.5,\allowbreak $ and 1, decreases to approximately $1.1a$ at $W/E_F=2$, and reaches approximately $0.9a$ at $W/E_F=4$. At the largest disorder, the corresponding values for $r_s=3,\allowbreak 8,\allowbreak $ and 16 are approximately $0.8a,\allowbreak 1.0a,\allowbreak $ and $1.0a$. These are nearest-neighbor coherence scales within the stated spectral convention. The clean $r_s=2$ density remains uniform in the selected search and is excluded from the clean-crystal comparison. Direct-only Hartree spectra and lengths at nonzero disorder describe an induced density response, not a disordering Hartree crystal.

The classical calculation resolves a larger range of separations. At $n_e=n_0$, the first-shell coherence extends beyond $5a$ through $c=1/16$. It decreases to approximately $2.7a$ at $c=1/8$, $1.2a$ at $c=1/4$, $0.9a$ at $c=1/2$, and $0.8a$ at $c=4$. The loss of the distant coordination shells and the conversion of sharp peaks into diffuse scattering therefore have a direct length-scale counterpart. Increasing disorder leaves only local packing around the pristine spacing.

\begin{figure*}[t]
    \centering
    \includegraphics[width=.95\textwidth]{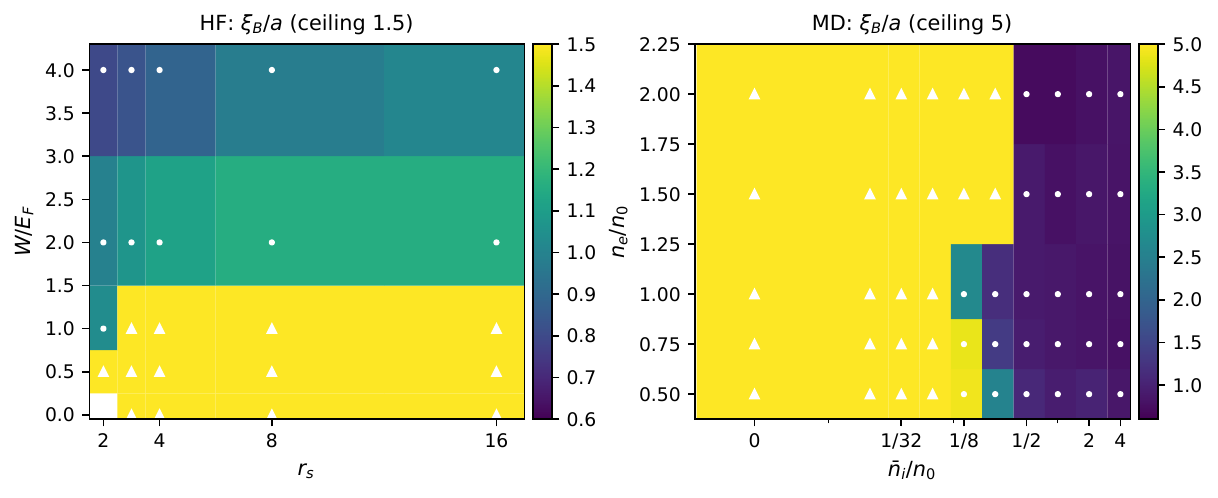}
    \caption{Translational coherence maps in the quantum $r_s$--$W/E_F$ plane and the classical impurity-density--electron-density plane. Colors show $\xi_B/a$ with display ceilings $1.5$ and 5; white triangles distinguish lower bounds from resolved points. The clean uniform HF point is blank because a density-wave coherence length is undefined there. Pixel boundaries are plotting cells and are not phase boundaries. The classical impurity axis is linear near zero and logarithmically compressed above $1/64$.}
    \label{fig:correlation_length_maps}
\end{figure*}

Figure~\ref{fig:correlation_length_maps} displays $\xi_B/a$ over both parameter planes: $r_s$ versus $W/E_F$ for HF and electron density versus impurity density for MD. Points above the accessible distance are marked as bounds rather than assigned an artificial plateau. These maps quantify a structural crossover. Their color changes and the joining lines in Fig.~\ref{fig:correlation_lengths_vs_disorder} do not locate a singularity. In the classical map, fixed well radius and depth compete differently with Coulomb forces as electron density changes; the resulting crossover need not be a universal function of the impurity-to-electron density ratio alone.

\begin{figure*}[t]
    \centering
    \includegraphics[width=.95\textwidth]{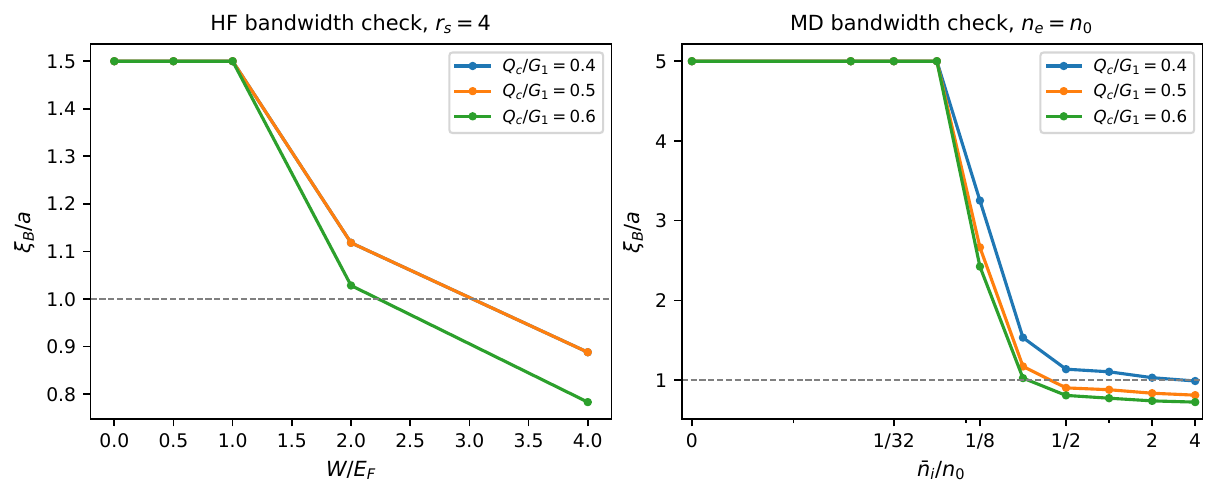}
    \caption{Sensitivity of the translational coherence length to carrier-patch bandwidth, at $r_s=4$ for HF and $q=1$ for MD. The three choices $Q_c/G_1=0.4,\allowbreak 0.5,\allowbreak 0.6$ preserve the decrease from extended to microscopic coherence but change its numerical endpoint. Censored values are shown at their measurement ceilings. At the main bandwidth, a flat diffuse patch alone has a first crossing at $0.71a$; lengths near that value indicate no resolved coherence beyond the local scale.}
    \label{fig:correlation_bandwidth_check}
\end{figure*}

\begin{figure*}[t]
    \centering
    \includegraphics[width=.95\textwidth]{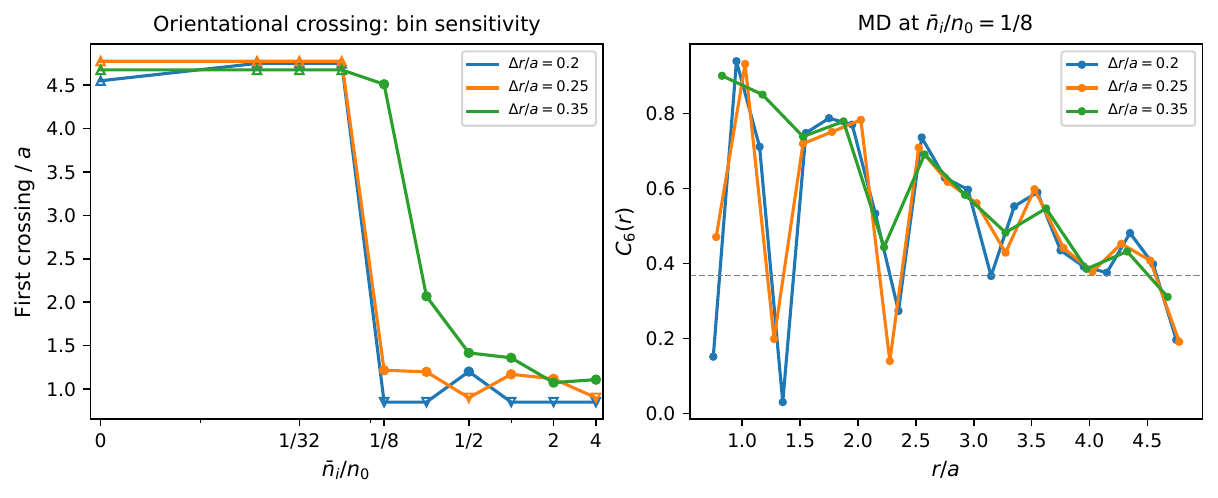}
    \caption{Sensitivity of the first bond-orientation crossing to radial bin width. Left: crossing values or bounds for widths $0.20a$, $0.25a$, and $0.35a$. Right: the corresponding $C_6(r)$ curves at $c=1/8$. Coordination-shell oscillations and low pair counts can move the first crossing substantially. These data are retained as a diagnostic of resolution; the paper uses the full orientational correlations and the translational length, rather than assigning a precise orientational decay length.}
    \label{fig:orientational_bin_sensitivity}
\end{figure*}

\subsection{Resolution, bandwidth, and bond orientation}\label{sec:resolution}

Figure~\ref{fig:correlation_bandwidth_check} tests the carrier-patch bandwidth. At $r_s=4$ and $W/E_F=4$, the HF estimate spans approximately $0.8a$--$0.9a$ for $\alpha=0.4$--0.6. At $n_e=n_0$ and $c=1/8$, the classical estimate spans approximately $2.4a$--$3.3a$; at $c=4$, it spans approximately $0.7a$--$1.0a$. Increasing the classical patch quadrature from 81 to 121 points per direction changes $\xi_B$ by less than 0.1\% at the latter two concentrations. The bandwidth convention, rather than numerical quadrature, controls the precision of these microscopic lengths.

An elementary calibration explains this limit. A flat intensity in a circular patch gives
\begin{equation}\label{eq:C_flat}
    C_{\mathrm{flat}}(r)=\frac{2J_1(Q_cr)}{Q_cr}.
\end{equation}
At $\alpha=0.5$, its first $e^{-1}$ crossing is $0.71a$. Thus a spectrum with no resolved narrow Bragg structure already produces a microscopic crossing through the imposed bandwidth. Once the measured length approaches this scale, its physical content is loss of coherence beyond approximately one lattice spacing. A precise sub-lattice correlation length is not resolved. The classical particle self term also contributes to this diffuse short-distance background. This floor is not small: its patch weight $\pi Q_c^2$ is $2\pi\alpha^2/\sqrt3\simeq0.91$ times that of a window-limited Bragg peak, $N_I(2\pi/\ell)^2$, at $\alpha=0.5$. Since $\min C_{\mathrm{flat}}\simeq-0.13$, $C_B$ then crosses $e^{-1}$ below $1.42a$ once the Bragg share of the patch weight falls below about 0.45, as it does for uncorrelated static displacements of root-mean-square magnitude $\gtrsim0.18a$ even with perfect long-range order. A classical $\xi_B$ of order $a$ thus also reflects lost Bragg intensity, as $p_0$ does for HF; the peak broadening in Figs.~\ref{fig:classical_snapshots_structure} and \ref{fig:disorder_structure_factors} and the orientational measures identify the loss of coherence. These effects do not account for the several-spacing coherence at weaker disorder, but they limit interpretation of its strong-disorder endpoint.

For an independent test of angular organization we evaluate
\begin{equation}\label{eq:C6}
    C_6(r)=\frac{\operatorname{Re}\langle\psi_6(i)\psi_6(j)^*\rangle_r}{\langle|\psi_6|^2\rangle_I},
\end{equation}
using the pair weights of Sec.~\ref{sec:order_measures} and annuli of width $0.25a$. The full curves in Fig.~\ref{fig:translational_orientational_correlations} distinguish the nearly coherent weak-disorder state from the strongly disordered state. Their first crossings, however, are sensitive to coordination-shell oscillations and sparsely populated distance bins. Figure~\ref{fig:orientational_bin_sensitivity} explicitly compares widths $0.20a$, $0.25a$, and $0.35a$. At $c=1/8$, the apparent first crossing ranges from below $0.9a$ to about $4.5a$. We therefore do not interpret this crossing as a separately resolved orientational decay length. The complete $C_6(r)$ curves, $\Psi_6$, and $P_6$ provide the orientational evidence; $\xi_B$ is the quantitative length used in the crossover maps. Individual orientational measures need not vary monotonically in a single disorder realization.

Independent numerical audits check the density normalization and Fourier conjugacy, the real-space/Fourier autocorrelation identity, and the classical structure factor against an explicit double particle sum. Their discrepancies are below $10^{-10}$. The angular Fourier integral agrees with the Bessel construction to the same tolerance. These checks validate the calculated observables; the finite observation range and bandwidth determine their physical resolution.

\section{Imry--Ma instability and Larkin length estimates}\label{sec:larkin}

\subsection{Why disorder produces a coherence crossover}\label{sec:imry_ma}

In an elastic description, the density contains harmonics $n_G\exp[iG\cdot(r-u(r))]$. A random potential couples directly to their local phases. A smooth distortion through displacement $u$ across a region of size $R$ costs transverse elastic energy of order $\mu u^2R^{d-2}$, while an uncorrelated random force can gain energy of order $\sqrt\Delta\,\allowbreak uR^{d/2}$. In two dimensions the latter outgrows the former with increasing $R$. This is the density-wave application of the Imry--Ma argument~\cite{imry1975random}~\cite{fukuyama1978dynamics}; the Larkin treatment resolves the corresponding weak-distortion scale~\cite{larkin1970effect}~\cite{larkin1979pinning}.

For an unscreened Coulomb crystal, the long-wavelength elastic energy can be written
\begin{equation}\label{eq:H_elastic}
    \begin{split}
        H_{\mathrm{el}}=\frac12\int\frac{d^2k}{(2\pi)^2}\bigl[&\mu k^2|u_T(k)|^2\\
        &+D_L(k)|u_L(k)|^2\bigr],
    \end{split}
\end{equation}
where $D_L(k)\simeq2\pi[e^2/(4\pi\epsilon)]n_e^2|k|+O(k^2)$. Coulomb forces stiffen longitudinal compression but leave the transverse $k^2$ shear mode. That soft mode is sufficient for the random-pinning instability. The relevant question is the distance over which a weakly distorted crystal remains coherent, not a positive threshold below which disorder leaves the pristine infinite crystal intact~\cite{chitra2005zero}.

This argument concerns loss of pristine long-range translational order. It does not equate weak elastic distortion with a high density of dislocations, or imply exponential decay with a unique length at every disorder strength. Here the term amorphous describes a pinned solid without pristine long-range crystalline organization; its weak-disorder end can retain large locally crystalline domains with no long range order whatsoever. The numerical crossover tracks how that residual organization decreases, ultimately reaching the nearest-neighbor scale.

\subsection{Definitions and infrared convention}\label{sec:larkin_definitions}

\begin{figure*}[t]
    \centering
    \includegraphics[width=.95\textwidth]{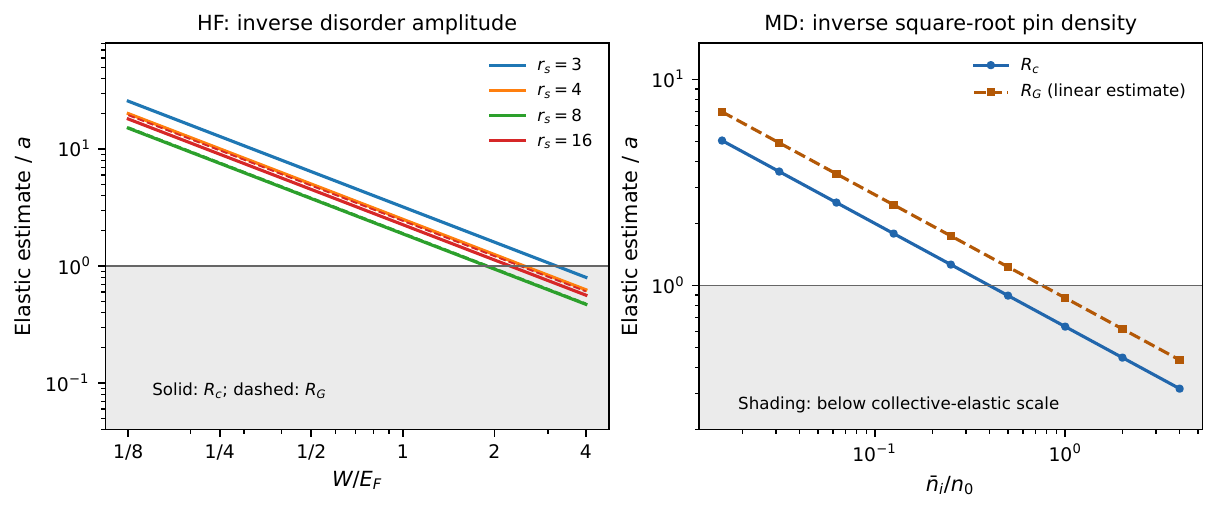}
    \caption{Qualitative Larkin estimates for both disorder models. Solid curves show $R_c$, defined by displacement reaching the force-correlation range; dashed curves show $R_G^{(L)}$, the linear Gaussian estimate for phase decorrelation. HF uses freshly calculated clean shear curvatures and the known grid-potential covariance. MD uses the pristine Coulomb shear modulus and the analytic force variance of the actual compact wells. The fixed infrared convention is specified in Eq.~\eqref{eq:infrared_convention}. Shading marks lengths below $a$, where collective elasticity breaks down. These estimates are distinct from the measured $\xi_B$ and involve no numerical disorder averaging.}
    \label{fig:larkin_length_estimates}
\end{figure*}

For an isotropic coarse-grained force density, define its disorder-law covariance by
\begin{equation}\label{eq:force_covariance}
    \overline{f_\alpha(r)f_\beta(r')}=\Delta\delta_{\alpha\beta}\delta^{(2)}(r-r').
\end{equation}
The overbar in this analytic model specifies the variance of the chosen disorder law. It does not represent an average of simulated samples; all numerical spectra and configurations remain those of the single quenched field. Linear transverse response gives
\begin{equation}\label{eq:BT}
    \begin{split}
        B_T(R)&=\overline{|u_T(R)-u_T(0)|^2}\\
        &\simeq\frac{\Delta}{\pi\mu^2}\int_{k_{\mathrm{min}}}^{k_{\mathrm{max}}}\frac{1-J_0(kR)}{k^3}\,dk.
    \end{split}
\end{equation}
With a fixed system infrared cutoff, the small-$k$ contribution has an $R^2\log(L/R)$ dependence in two dimensions. To give reproducible qualitative domain estimates, we instead use $k_{\mathrm{min}}=1/R$ and extend the upper limit to infinity when $R\gg a$. This convention gives
\begin{equation}\label{eq:infrared_convention}
    \begin{split}
        J&=\int_1^\infty\frac{1-J_0(x)}{x^3}\,dx,\\
        B_{\mathrm{dom}}(R)&=\frac{\Delta J}{\pi\mu^2}R^2,\qquad A_L=\sqrt{\pi/J}\simeq3.31.
    \end{split}
\end{equation}
Let $b$ be the displacement over which the local pinning force changes appreciably, defined through $C_f(u)/C_f(0)=1-u^2/b^2+\cdots$. The Larkin length $R_c$ is where the linear displacement reaches $b$. We also quote the linear estimate $R_G^{(L)}$ where an isotropic Gaussian density-wave correlator $\exp[-G_1^2 B(R)/4]$ would fall to $e^{-1}$:
\begin{equation}\label{eq:larkin_lengths}
    R_c=A_L\frac{\mu b}{\sqrt\Delta},\qquad R_G^{(L)}=A_L\frac{2\mu}{G_1\sqrt\Delta}.
\end{equation}
These are distinct from the measured $\xi_B$. In particular, $R_G^{(L)}$ can extend beyond the regime where the random-force linearization remains valid. The infrared convention and neglected nonlinear pinning fix only an order-unity prefactor; the disorder dependence is the useful qualitative result. Collective elasticity ceases to apply once the estimated length approaches or falls below $a$. Figure~\ref{fig:larkin_length_estimates} shades this region rather than treating the extrapolation as a prediction of fractional-lattice domains.

\subsection{Quantum Hartree--Fock estimate}\label{sec:larkin_hf}

We calculate the clean HF shear stiffness from relaxed, area-preserving strains. Apply $F(s)=\left(\begin{smallmatrix}1&s\\0&1\end{smallmatrix}\right)$ to the real-space lattice and transform reciprocal vectors with $F^{-T}$. The kinetic energy, direct field, and every exchange quadrature cell are updated consistently. At each strain we solve the orbitals and occupations again, using the same basis and temperature as in the principal scan. If $f(s)$ is the free energy per electron, the shear coefficient is
\begin{equation}\label{eq:shear_HF}
    \mu_{\mathrm{HF}}=n_e\lim_{s\to0}\frac{f(s)+f(-s)-2f(0)}{s^2}.
\end{equation}
Twenty clean SCF calculations cover $r_s=3,\allowbreak 4,\allowbreak 8,\allowbreak 16$ at $s=0,\allowbreak \pm0.01,\allowbreak \pm0.02$. All fixed-point residuals are below $10^{-7}$. The two strain steps give curvatures agreeing within 0.06\%. In units $H^*/(a_B^*)^2$, the resulting shear coefficients are approximately $3.07\times10^{-3}$, $1.07\times10^{-3}$, $9.72\times10^{-5}$, and $1.18\times10^{-5}$, respectively. The symmetric difference removes the linear stress of the discrete calculation. These are stiffnesses of the chosen finite-basis HF branch; strain-step agreement does not establish basis or mesh convergence of the elastic constants.

Write $\mathcal A_c=9\pi(r_sa_B^*)^2$ for the physical cell area and $n_G=\rho_G/\mathcal A_c$ for the clean density amplitudes. The 1024 independent grid variables of the HF disorder model have physical spectral covariance $C_U=\mathcal A_c W^2/1024$ at represented nonzero modes. Expanding their coupling to the clean crystal gives the isotropic force variance and curvature range
\begin{equation}\label{eq:disorder_strength_HF}
    \begin{split}
        \Delta_{\mathrm{HF}}&=\frac{C_U}{2}\sum_{G\ne0}G^2|n_G|^2,\\
        b_{\mathrm{HF}}&=2\left[\frac{\sum_G G^2|n_G|^2}{\sum_G G^4|n_G|^2}\right]^{1/2}.
    \end{split}
\end{equation}
All vectors in Eq.~\eqref{eq:disorder_strength_HF} have physical inverse-length units. The displacement range is approximately $0.26a$--$0.28a$ across the four crystalline clean states. A first-shell-only density gives $b_{\mathrm{HF}}=2/G_1\simeq0.28a$ and hence $R_G^{(L)}=R_c$; the computed $R_G^{(L)}$ exceeds $R_c^{\mathrm{HF}}$ by at most 8\%, so the two HF curves in Fig.~\ref{fig:larkin_length_estimates} nearly coincide. With $\eta=W/E_F$, the resulting Larkin estimates are
\begin{equation}\label{eq:Rc_HF}
    \frac{R_c^{\mathrm{HF}}}a\simeq\frac{1}{\eta}\begin{cases}3.2,&r_s=3,\\2.5,&r_s=4,\\1.9,&r_s=8,\\2.3,&r_s=16.\end{cases}
\end{equation}
At fixed $r_s$, the weak-pinning scaling is therefore $R_c\propto W^{-1}$. The estimates use the clean stiffness and density amplitudes; strong disorder changes both and invalidates that extrapolation. At $r_s=4$ and $\eta=0.5$, the estimate is about $5a$, already larger than the simulated quantum cell. At $\eta=4$ the formal value is about $0.6a$, identifying breakdown of a collective crystal description rather than resolving a sub-lattice elastic domain. This is consistent with the measured intra-cell coherence becoming microscopic. No crystal Larkin length is assigned to clean direct-only Hartree, whose density is uniform.

\subsection{Classical molecular-dynamics estimate}\label{sec:larkin_classical}

The pristine triangular Coulomb lattice has shear modulus~\cite{bonsall1977some}
\begin{equation}\label{eq:shear_classical}
    \mu_{\mathrm{cl}}=0.245065\,\frac{e^2}{4\pi\epsilon}\,n_e^{3/2}.
\end{equation}
At $q=1$, the compact wells satisfy $2R_p<a$, so independent Poisson impurities give uncorrelated forces at distinct pristine lattice sites. For the force derived from Eq.~\eqref{eq:pin_potential}, direct radial integration yields
\begin{equation}\label{eq:disorder_strength_classical}
    \begin{split}
        \int d^2r\,|\nabla v_p|^2&=\frac{4\pi D^2}{3},\\
        \Delta_{\mathrm{cl}}&=\frac{2\pi}{3}n_e\bar n_i D^2,\qquad b_{\mathrm{cl}}=R_p/2.
    \end{split}
\end{equation}
The last relation follows from the curvature of the force autocorrelation, not from assigning the well radius itself as a displacement threshold. Substituting the actual $D=0.1E_0$ and $R_p=0.4a$ at $q=1$ gives
\begin{equation}\label{eq:Rc_classical}
    \begin{split}
        \frac{R_c^{\mathrm{cl}}}a&\simeq\frac{0.63}{\sqrt c},\qquad\frac{R_G^{(L)}}a\simeq\frac{0.87}{\sqrt c},\\
        c&=\bar n_i/n_0.
    \end{split}
\end{equation}
Thus increasing impurity concentration reduces the weak-pinning length as $\bar n_i^{-1/2}$ at fixed density and well parameters. Increasing well depth instead would give $R_c\propto D^{-1}$. At $c=1/8$ the estimates are $R_c\simeq1.8a$ and $R_G^{(L)}\simeq2.5a$, compared with the measured $\xi_B\simeq2.7a$. This order-of-magnitude agreement uses no fitted parameter and is not a validation of the prefactor; a Larkin disk holds only about 1.5 wells at every $c$, so $R_c$ simply tracks the mean well spacing, $2.6a$ at $c=1/8$. At $c=4$, the formal elastic lengths are below $a$ and are no longer quantitatively meaningful, while the measured spectrum has only microscopic coherence. The estimate is restricted to $q=1$; at sufficiently high electron density, neighboring-site force correlations must be retained because $2R_p<a$ no longer holds.

Both microscopic descriptions therefore give the same elastic tendency: stronger random forces shorten the distance over which a pristine lattice can resist distortion. The numerical spectra follow that tendency into the regime where a collective elastic domain is no longer resolved. The HF amplitude scan and the classical concentration scan have different power laws because their random-force variances scale as $W^2$ and $\bar n_i D^2$, respectively.

\section{Discussion}\label{sec:discussion}

\subsection{Numerical scope and resolution}\label{sec:scope}

The quantum and classical studies consistently show loss of crystalline organization with increasing random pinning. The evidence is a transfer of static spectral weight, shortening of the first-shell density coherence, and, in the particle calculation, degradation of bond order and coordination. The disorder-dependent lengths supplement the threshold maps with a continuous structural measure. Once $\xi_B$ reaches order $a$, there is no resolved coherent crystal domain beyond local packing within that convention.

The interpretation of weak disorder requires separating observation scale from long-range order. A finite image can remain crystal-like because the disorder-induced distortion scale exceeds its dimensions. Random impurities nevertheless remove the pristine translationally invariant arrangement. The Imry--Ma and Larkin reasoning describes the long-scale instability; the present small quantum cell does not itself simulate that limit. In the broader noncrystalline sense specified in Sec.~\ref{sec:larkin}, disorder produces an amorphous pinned solid whose residual crystalline character depends on disorder strength. It need not be locally defect-rich at arbitrarily weak pinning. We identify no finite-disorder crystal-to-amorphous phase transition in these maps; their boundaries mark chosen levels of residual order.

The HF calculation periodically repeats a nine-electron cell and resolves only a sparse set of density-wave offsets. Its $\xi_B$ is explicitly an intra-cell coherence diagnostic, while its Larkin estimate concerns weak random pinning of the clean elastic state. The classical $N=288$ calculation resolves distances through $5a$ inside an open observation square. Its microscopic length depends on spectral bandwidth, and the first crossing of its oscillatory bond-orientation function is not a stable decay-length estimator. These concrete restrictions preclude extracting asymptotic exponents, an exact sub-lattice coherence length, or a dislocation transition from the present data.

Both scans use one disorder field; no disorder average or thermodynamic-limit extrapolation is performed. Changing the HF Bragg-fraction threshold from 0.5 to 0.4 shifts three brackets by one disorder interval. The classical structural criteria are fixed at $\Psi_6\geq0.5$ and $P_6\geq0.85$. Longer heating trajectories expose the dilute-disorder sampling dependence while preserving the increase in the strong-pinning cage-escape scale. The $N=572$ check concerns the clean system and gives a thermal crossover scale of roughly $\theta=0.005$--0.010. All disordered heating results retain $N=288$, damping $\gamma=0.5$, and the specified observation lag. They characterize thermal rearrangement under that protocol, rather than a bulk equilibrium melting line through an amorphous regime.

\subsection{Method comparison}\label{sec:method_comparison}

HF occupation-temperature checks and MD heating answer different questions. The former test the stability of the self-consistent density against occupation smearing; the latter measure particle rearrangement. Vu and Das Sarma found disorder enhancement of an effective melting temperature in exact-diagonalization calculations using localization and conductance diagnostics~\cite{vu2022thermal}. Our MD results identify a classical route to a higher mobility crossover temperature: stronger trapping persists after crystalline coherence has been lost.

HF treats fermionic exchange in a periodic cell; MD treats interacting classical particles in an open square. Their disorder models and boundary conditions differ. The shared result is the separation of density modulation, crystalline coherence, and mobility, not a numerical identification of their disorder or temperature scales.

\subsection{Relation to the disputed MoSe$_2$ Wigner-crystal interpretation}\label{sec:mose2}

The imaging studies~\cite{xiang2025imaging,ge2026visualizing} and their interpretation~\cite{babbar2026wigner} place real-space density patterns and Fourier peaks at the center of the Wigner-solid discussion. Our calculations show concretely why the amplitude of a density pattern is insufficient to establish crystalline order. In HF, density contrast grows while the fraction of weight at the pristine reciprocal vectors falls. In MD, sharp peaks become diffuse and orientational correlations weaken, yet rearrangements remain slow. Strong spatial modulation and strong pinning therefore coexist with amorphous structure.

This distinction is directly relevant to interpreting finite images of electron solids. The spatial extent and scaling of crystalline correlations test order; mobility and transport test charge rearrangement and conduction. Neither density contrast nor an insulating response alone identifies both properties. Applying the present calculation quantitatively to MoSe$_2$ requires its screened interaction, disorder statistics, spin and valley degrees of freedom, and the relation between the calculated density and the imaging signal. The calculation addresses structural disordering and classical trapping; it does not compute quantum localization.

\section{Conclusions}\label{sec:conclusion}

We have studied impurity-induced disordering of a two-dimensional Wigner solid in two independent microscopic descriptions: quantum Hartree--Fock theory and classical molecular dynamics. The two studies address the same physics and reach the same structural conclusion. Random pinning progressively removes crystalline coherence, while the electron density or particle positions remain strongly inhomogeneous. Disorder controls how much crystalline organization survives within a pinned noncrystalline solid.

In the quantum study, the Hartree/HF comparison isolates exchange: the clean direct-only Hartree density is uniform, whereas the selected HF states at $r_s=3,\allowbreak 4,\allowbreak 8,\allowbreak $ and 16 are crystalline. Increasing disorder moves density weight away from the pristine reciprocal vectors. At $r_s=4$, the operational intra-cell coherence length changes from an unresolved value above $1.5a$ at weak disorder to approximately $1.1a$ and $0.9a$ at $W/E_F=2$ and 4. The density contrast grows during this loss of crystalline organization. Large modulation therefore cannot serve as a proxy for crystal order.

In the classical study, the structure factor evolves from six sharp peaks into diffuse scattering, and both positional and orientational correlations weaken. At $n_e=n_0$, the measured translational coherence decreases from more than $5a$ at weak pinning to approximately $2.7a$, $1.2a$, and $0.8a$ at impurity concentrations $c=1/8$, $1/4$, and 4. The two-density map extends this result across electron density. The strong-disorder endpoint is a mechanically stable amorphous configuration with only local packing order; the bandwidth checks establish an order-$a$ scale rather than a precise distance below one lattice constant.

The Imry--Ma and Larkin description supplies the common long-distance interpretation. Arbitrarily weak random pinning destabilizes pristine translational order on a sufficiently large scale, while increasing disorder shortens the distortion scale. Our explicit weak-pinning estimates give $R_c\propto W^{-1}$ for the HF amplitude scan and $R_c\propto\bar n_i^{-1/2}$ for the classical concentration scan; the latter is also the scaling of the mean well spacing, since each classical Larkin disk holds only about 1.5 wells. The numerical diagrams consequently describe varying degrees of noncrystallinity, not a finite-disorder boundary at which a pristine crystal suddenly becomes amorphous. Weakly pinned domains can still look crystalline over a finite observation window.

Classical heating reveals a second, distinct consequence of impurities. The primary cage-escape bracket rises from $k_BT/E_0=0.005$--0.007 in the clean system to 0.020--0.030 at $c=4$, where crystalline order has already disappeared at zero temperature. Stronger disorder therefore stabilizes an amorphous pinned assembly against rearrangement while degrading its spatial order. Crystalline coherence and resistance to motion are distinct properties: disorder suppresses the former while strengthening the latter.

We conclude that a claim that extrinsic disorder enhances Wigner crystallization and lowers the critical $r_s$ for crystallization is misleading---what happens is that disorder may enhance some short range order, but the long range order characteristic of a crystal is always suppressed by disorder, the system becomes an amorphous solid. Indeed, if increasing disorder enhances Wigner crystallization, one could arbitrarily lower the critical $r_s$ (increase the critical density) simply by adding impurities to the system, which makes little sense because high impurity density must always lead to an amorphous structure.

\section*{Acknowledgments}
S.D.S. is supported by the Laboratory for Physical Sciences through the Condensed Matter Theory Center at the University of Maryland.
H.P. is supported by startup funds at the University of Florida.
The authors acknowledge the usge of GPT-6 Astra to perform the numerical work and plot the figures, and Claude Fable 5.1 to double-check the numerical work and the figures.

\bibliography{WignerDisorder,Paper_WignerDisorder}
\end{document}